\documentclass[journal]{IEEEtran}
\pdfoutput=1
\usepackage{amsmath} 
\usepackage[english]{babel}

\usepackage{graphicx}
\usepackage{enumerate}
\usepackage{graphics}
\usepackage{enumitem}
\usepackage{amssymb}
\usepackage{float}
\usepackage{rotating}
\usepackage{comment}
\usepackage{copyrightbox}
\usepackage{rotating}
\usepackage{booktabs}
\usepackage{colortbl}
\usepackage{multicol}
\usepackage{multirow}

\usepackage{cite}
\usepackage{hyperref}
\usepackage{dblfloatfix}
\usepackage{amsthm}
\usepackage{mathrsfs} 
\newtheorem{proposition}{Proposition}
\newtheorem{assumption}{Assumption}

\newcommand{\proba}{\text{p}}
\DeclareMathOperator*{\argmin}{\text{arg min}}

\newcommand{\V}[1]{\boldsymbol{#1}}
\newcommand{\M}[1]{\mathbf{#1}}

\newcommand{\T}{^\dagger}

\newcommand{\centered}[1]{\dot{#1}}
\newcommand{\whitened}[1]{\mathring{#1}}

\newcommand{\diag}{\mathrm{diag}}

\begin{document}
\title{
Multi-temporal speckle reduction with
self-supervised deep neural networks
}

\author{Inès~Meraoumia, Emanuele~Dalsasso, Lo{\"i}c~Denis, Rémy Abergel, and Florence~Tupin%
\thanks{I. Meraoumia, E. Dalsasso and F. Tupin are with LTCI, Télécom Paris, Institut Polytechnique de Paris, Palaiseau, France, e-mail: forename.name@telecom-paris.fr.}%
\thanks{L. Denis is with Univ Lyon, UJM-Saint-Etienne, CNRS, Institut d Optique Graduate School, Laboratoire Hubert Curien UMR 5516, F-42023, SAINT-ETIENNE, France, e-mail: loic.denis@univ-st-etienne.fr.}%
\thanks{R. Abergel is with the laboratoire MAP5, UMR CNRS 8145, Université Paris Cité, France, e-mail: remy.abergel@parisdescartes.fr.}%
}
\maketitle
\begin{abstract}
Speckle filtering is generally a prerequisite to the analysis of synthetic aperture radar (SAR) images. Tremendous progress has been achieved in the domain of single-image despeckling. Latest techniques rely on deep neural networks to restore the various structures and textures peculiar to SAR images. The availability of time series of SAR images offers the possibility of improving speckle filtering by combining different speckle realizations over the same area.

The supervised training of deep neural networks requires ground-truth speckle-free images. Such images can only be obtained indirectly through some form of averaging, by spatial or temporal integration, and are imperfect. Given the potential of very high quality restoration reachable by multi-temporal speckle filtering, the limitations of ground-truth images need to be circumvented. We extend a recent self-supervised training strategy for single-look complex SAR images, called MERLIN, to the case of multi-temporal filtering. This requires modeling the sources of statistical dependencies in the spatial and temporal dimensions as well as between the real and imaginary components of the complex amplitudes.

Quantitative analysis on datasets with simulated speckle indicates a clear improvement of speckle reduction when additional SAR images are included. Our method is then applied to stacks of TerraSAR-X images and shown to outperform competing multi-temporal speckle filtering approaches.

 The code of the trained models is made freely available at \url{https://gitlab.telecom-paris.fr/ring/multi-temporal-merlin/}.

\end{abstract}

\IEEEpeerreviewmaketitle

\section{Introduction}

Earth Observation requires diverse information that can be captured with complementary remote sensing systems.
Synthetic Aperture Radar (SAR) is an active sensor widely used
in applications ranging from ocean and forest monitoring, land use and human activity monitoring, to the estimation of digital elevation models \cite{moreira2013tutorial}.

However, interpreting SAR images is particularly challenging because of the presence of strong fluctuations in the back-scattered intensities: 
due to the coherent sum of the contributions of all scatterers located within the same resolution cell, constructive or destructive interferences occur, leading to the so-called \emph{speckle} phenomenon.
SAR image analysis is greatly simplified when speckle fluctuations are reduced in a pre-processing step.

Speckle reduction has been tackled by various approaches, from methods based on the selection of pixels with similar intensities \cite{lee1983digital}, to techniques based on wavelet decompositions \cite{argenti2013tutorial} or non-local filtering \cite{deledalle2014exploiting}, and, more recently, significant progress was achieved with deep neural networks \cite{zhu2021deep,fracastoro2021deep,rasti2021image}.

Training a neural network for speckle reduction requires the definition of a loss function which reflects the performance of the network on the training set. In a \emph{supervised training} setting, this loss function characterizes the proximity of the network prediction to a ground truth image: the speckle-free image corresponding to the ideal output. Building a training set with matching pairs of speckle-free and corrupted SAR images is difficult. Starting from a corrupted SAR image, creating the corresponding speckle-free ground truth has no ideal solution (in fact, this is our ultimate goal). Speckle-free images can be obtained either by another modality (e.g., optical remote sensing, natural images) or by computing the temporal mean of a long time series of SAR images  \cite{vitale2022analysis}. Once a speckle-free image is selected, a corrupted version can be produced by drawing samples from a theoretical distribution of speckle. To prevent any domain shift between the training and testing phases, speckle simulation has to accurately capture the actual speckle fluctuations observed in SAR images, in particular its spatial correlations. An alternative is to define a \emph{self-supervised training} loss, i.e., a loss function relating the network estimation to other observations. SAR2SAR \cite{dalsasso2021sar2sar} extends to speckle reduction the Noise2Noise principle \cite{lehtinen2018noise2noise}: the denoised image should be close, on average, to other independent noisy observations of the same scene. Because of changes occurring between image acquisitions, special care must be taken to compensate these changes. Single-image self-supervised training is also possible. Speckle2Void \cite{molini2020speckle2void} follows the blind spot methodology introduced in \cite{laine2019high} that excludes the central pixel from the network estimation in order to drive the training step (by minimizing the statistical distance between the speckled central pixel and the despeckled network prediction based solely on the surrounding area). Rather than spatially splitting the input image into blind spots and surrounding areas, MERLIN \cite{MERLIN} splits the Single Look Complex (SLC) input image into the real and imaginary parts to define the self-supervised loss.

Time series offer more information to reduce speckle fluctuations than a single SAR image.
Multi-temporal averaging can be largely improved by compensating for changes, as proposed in Quegan filter \cite{Quegan}, up to a limit depending on the length of the time series and the quality of single-image restorations used for change suppression. 
Successful single-image despeckling techniques have been extended to multi-temporal data: SAR-BM3D \cite{parrilli2011nonlocal}, based on collaborative filtering of blocks of similar patches, also considers patches located at other dates in the multi-temporal extension \cite{block_matching}; the two-step multi-temporal non-local means
\cite{su2014two} perform weighted averages along the temporal or spatial dimensions based on patch similarities \cite{deledalle2009iterative}.
RABASAR \cite{RABASAR} proposes to compute first a "super-image" by temporally multi-looking the image stack (this super-image has almost no residual speckle fluctuations) and then process ratio images in which only speckle and changes with respect to the super-image are remaining. The content of these ratio images is largely simplified and thus easier to restore. The final despeckled images are obtained after multiplication by the super-image. To despeckle the ratio image, a deep neural network such as SAR2SAR can be used \cite{RABASAR-SAR2SAR}. A drawback of ratio-based processing is that the lowest-contrasted structures present either in the speckled image or in the super-image might be improperly restored, leading to the suppression of these details or the apparition of a "ghost" structure leaking from the super-image.

\emph{Our contributions:} We show how a deep neural network can be trained end-to-end to produce a despeckled image from a time series of co-registered SAR images. This is made possible by the use of a self-supervised loss function \cite{MERLIN}, bypassing the impossibility to access to high-quality ground truth images. Compared to simpler strategies based only on a single date enriched by a higher signal-to-noise multi-temporal average (a super-image) \cite{RABASAR}, we feed the network with all available dates. This leaves all freedom to the network to perform optimal temporal combinations, leading to improved restorations even when only a few additional images are included.

Our method is grounded on a generative model of speckle that accounts for fully-developed speckle areas, the presence of dominant scatterers due to man-made structures, interferometric coherence (both temporal and geometrical decorrelation phenomena), and the spatial correlations induced by the SAR transfer function.

The theoretical framework of the method is developed in section \ref{sec:proposed_app}.
A numerical study is then performed on data with simulated speckle to characterize the performance of the method. 
The approach is then tested on stacks of TerraSAR-X Stripmap images.

\section{Proposed approach: multi-temporal MERLIN}
\label{sec:proposed_app}

To derive a self-supervised training strategy in the context of multi-temporal filtering, we start by building a generative model of speckle in paragraph \ref{sec:genemodel}. We then discuss in paragraph \ref{sec:condindep} conditions under which a component of the reference date, statistically independent from the rest of the data, can be set aside in order to drive the training of deep neural networks. In section \ref{sec:training} we describe our unsupervised training strategy and the network architecture choices.

\begin{table}[t]
    \centering
    \begin{tabular}{@{}c@{\hspace*{1.5ex}}c@{\hspace*{1.5ex}}l@{}}
        \toprule
        \multicolumn{3}{l}{\hspace*{-1ex}\emph{Vector notations:}}\\
        $\V z$ & $\mathbb{C}^{TN}$ & representation of a stack of $T$ $N$-pixels images\\
        $\V z(\cdot,k)$ & $\mathbb{C}^{T}$ & vector of values at pixel $k$ \\
        $\V z(t,\cdot)$ & $\mathbb{C}^{N}$ & $t$-th image of the stack \\
        $\V z_t$ & $\mathbb{C}^{N}$ & $t$-th image of the stack (compact notation) \\
        $\V z_{\text{ref}}$ & $\mathbb{C}^{N}$ & image at date $t_{\text{ref}}$, the date to restore \\        
        \midrule
       \multicolumn{3}{l}{\hspace*{-1ex}\emph{Scene parameters:}}\\
        $\V d$ & $\mathbb{C}^{TN}$ & dominant scatterers \\
        $\V r$ & $\mathbb{R}_{+*}^{TN}$ & reflectivities of speckled areas \\
        \midrule
        \multicolumn{3}{l}{\hspace*{-1ex}\emph{Speckle field:}}\\
        $\V \epsilon$ & $\mathbb{C}^{TN}$ & uncorrelated speckle\\
        $\M \Gamma_k$ & $\mathbb{C}^{T\times T}$ & speckle coherence matrix at pixel $k$\\
        $\M L_k$ & $\mathbb{C}^{T\times T}$ & correlating operator such that $\M L_k\M L_k\T=\M \Gamma_k$\\
        $\M L$ & $\mathbb{C}^{TN\times TN}$ & correlating operator for the full stack\\
        \midrule
        \multicolumn{3}{l}{\hspace*{-1ex}\emph{Complex amplitudes on the radar antenna:}}\\
        $\V s$& $\mathbb{C}^{TN}$& complex amplitude of the speckled component\\
        $\V z$& $\mathbb{C}^{TN}$& resultant complex amplitude: $\V z = \V s+\V d$ \\
        $\tilde{\V z}$& $\mathbb{C}^{TN}$& complex amplitude including SAR system effects\\ 
        \midrule
        \multicolumn{3}{l}{\hspace*{-1ex}\emph{Acquisition specific parameters:}}\\
        $\V \varphi_t$& $\mathbb{C}^{N}$& atmospheric, topographic, and displacement\\
        && phase effects at each pixel of the $t$-th image\\
        $\V \psi_t$& $\mathbb{C}^{N}$ & phase ramp corresponding to the spectrum shift\\ &&due to angular discrepancies\\
        $\M Q$ & $\mathbb{C}^{N\times N}$ & SAR response (spectral apodization and 0-padding)\\
        $\M H_t$ & $\mathbb{C}^{N\times N}$ & SAR response (spectral apodization, 0-padding+shift)\\
        \midrule
        \multicolumn{3}{l}{\hspace*{-1ex}\emph{Pre-processing step to enforce statistic independence:}}\\
        $\centered{\V z}$& $\mathbb{C}^{TN}$& complex amplitudes with recentered power spectrum\\
        $\gamma_{ij}(k)$ & $\mathbb{C}$ & complex correlation coefficient (i.e., coherence)\\
        &&between $\centered{\V z}(t_i,k)$ and $\centered{\V z}(t_j,k)$\\
        $\M W_k$ & $\mathbb{C}^{2\times 2}$ & whitening matrix at pixel $k$\\
        $\M W$ & $\mathbb{C}^{2N\times 2N}$ & whitening operator for a pair of images\\
        $\whitened{\V z}$& $\mathbb{C}^{TN}$& complex amplitudes after whitening\\
        \midrule
        \multicolumn{3}{l}{\hspace*{-1ex}\emph{Self-supervised training:}}\\
        $\whitened{\V a}_{\text{ref}}$& $\mathbb{C}^{N}$& real part of pre-processed image at date $t_{\text{ref}}$\\
        $\whitened{\V b}_{\text{ref}}$& $\mathbb{C}^{N}$& imaginary part of pre-processed image at date $t_{\text{ref}}$\\
        $\mathcal{L}_{\text{MERLIN}}$&& self-supervised loss function\\
        $\tilde{\V r}_{\text{ref}}$ & $\mathbb{R}_{+*}^{N}$ & low-pass filtered reflectivities at date $t_{\text{ref}}$\\
        $\centered{\V d}_{\text{ref}}$ & $\mathbb{C}^{N}$ & low-pass filtered dominant scatterers at date $t_{\text{ref}}$\\
    \bottomrule
    \end{tabular}\\[1ex]
    \caption{Main notations and corresponding dimensions.}
    \label{tab:notations}
\end{table}

\subsection{Generative speckle model of multi-temporal SLC stacks}
\label{sec:genemodel}
The ability to partition the data into two mutually independent sets is central to our self-supervised training strategy. It is thus necessary to model the different sources of speckle correlations arising in multi-pass SAR imaging. If the images are acquired in interferometric conditions, then the speckle remains partially coherent from one pass to the next. Otherwise, the speckle is fully decorrelated and multi-temporal filtering can be very effective.

\begin{figure*}[t]
    \centering
    \includegraphics[width=\textwidth]{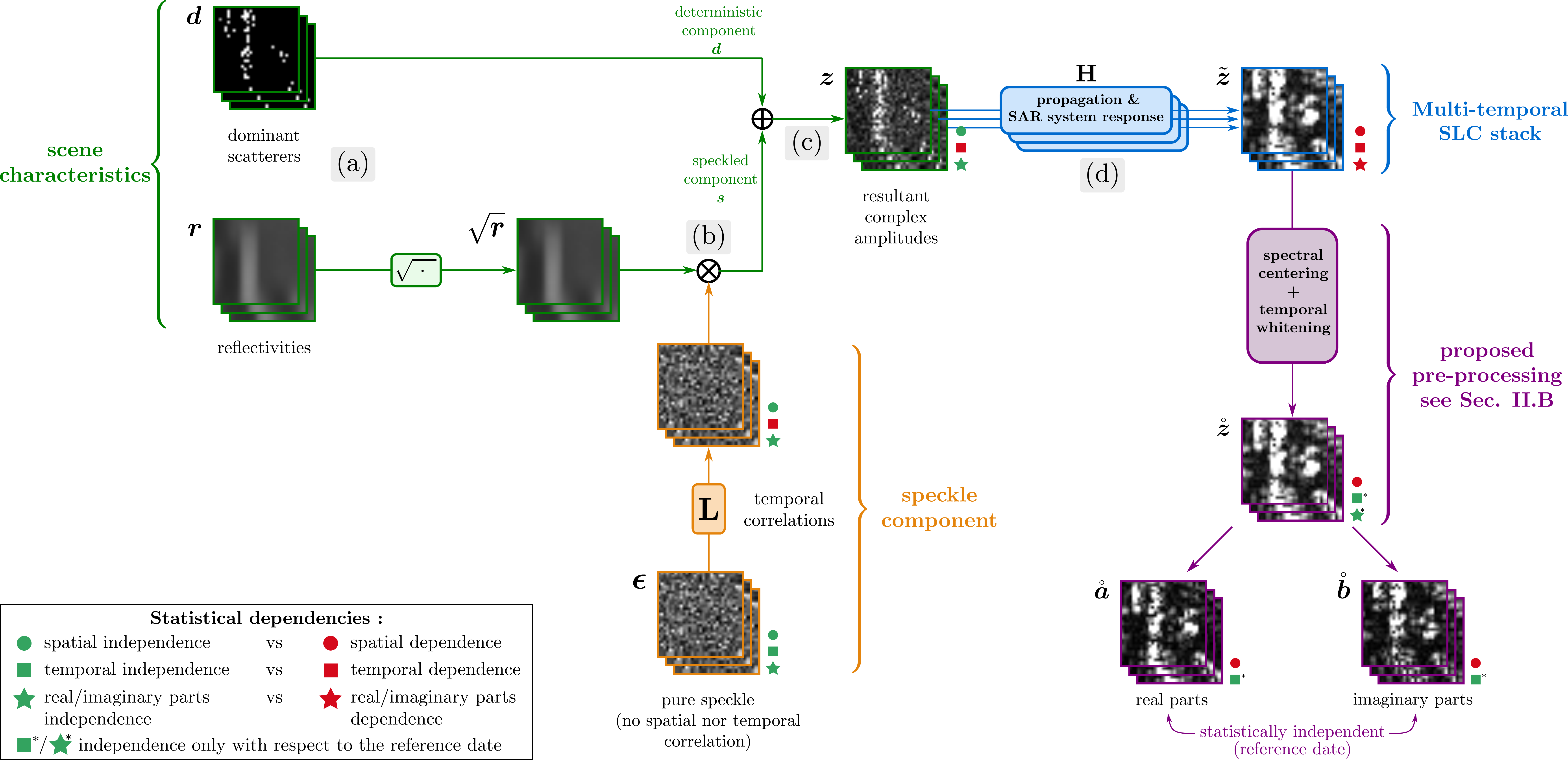}
    \caption{Generative model of speckle in multi-temporal SLC stacks of SAR images.}
    \label{fig:GM}
\end{figure*}

We consider here a more general speckle model than in \cite{MERLIN}, to account for the mix present in SAR images between (i) areas that follow Goodman's fully developed model (coherent summation of many similar elementary phasors), composed of rough surfaces and scattering volumes, and (ii) regions where the complex amplitude is mainly defined by the magnitude and phase of dominant scatterers. To include both phenomena, we model a stack $\V z\in \mathbb{C}^{TN}$ of $T$ SLC SAR images, each with $N$ pixels, as the superimposition of two components: a \emph{speckle component} 
$\V s \in \mathbb{C}^{TN}$, driven by a reflectivity map $\V r\in\mathbb{R}_{+*}^{TN}$, 
and the \emph{dominant scatterers component} $\V d\in \mathbb{C}^{TN}$, see Figure \ref{fig:GM}(a).

In the following, the multi-temporal stacks will be represented in the form of a column vector (e.g., $\V z\in\mathbb{C}^{TN}$), by concatenation of the $T$ images, and both the image at date $t$ (noted $\V z(t,\cdot)\in\mathbb{C}^{N}$, or $\V z_t$ in compact form) and the vector of complex amplitudes at pixel $k$ for all dates (noted $\V z(\cdot,k)\in\mathbb{C}^{T}$) will be considered. A permutation matrix $\M \Pi$ can be applied to transform the vector $\V z$ from an ordering according to a scan of all pixels for each date, one date after another, to an ordering according to a scan of all dates for a given pixel, before moving to the next pixel:
\begin{align}
\M{\Pi}\V z=
    \M{\Pi}
    \begin{pmatrix}
    \V z(t_1,\cdot)\\
    \vdots\\
    \V z(t_T,\cdot)
    \end{pmatrix} =
    \begin{pmatrix}
    \V z(\cdot,k_1)\\
    \vdots\\
    \V z(\cdot,k_N)
    \end{pmatrix}\,.
\end{align}
Table \ref{tab:notations} summarizes the main notations used in the paper.

According to Goodman's model \cite{goodman2007speckle}, the \emph{speckle component} $\V s(\cdot,k)\in\mathbb{C}^T$  at pixel $k$ follows a complex circular Gaussian distribution $\mathcal{N}_c(\M \Sigma_k)$ defined by
\begin{align}
    \proba(\V s(\cdot,k)|\M \Sigma_k) = \frac{1}{\pi^T
    \text{det}(\M \Sigma_k)}\exp\left[-\V s(\cdot,k)\T\,\M \Sigma_k^{-1}\,\V s(\cdot,k)\right],
\end{align}
where $\cdot\T$ denotes the conjugate transpose;  $\M \Sigma_k$ is the speckle covariance matrix at pixel $k$, which can be factored as $\M \Sigma_k=\text{diag}(\sqrt{\V r(\cdot,k)})\M \Gamma_k\text{diag}(\sqrt{\V r(\cdot,k)})$ with $\M \Gamma_k$ the coherence matrix (the entries verify $|\M \Gamma_k(t_i,t_j)|\leq 1$ for all $t_i$ and $t_j$ and $\M \Gamma_k(t,t)=1$ for all $t$), $\V r\in\mathbb{R}_{+*}^{TN}$ is the vector of reflectivities, and the square root is applied entry-wise. The coherence matrices characterize how the temporal evolution of the scene decorrelates the speckle.
Starting from a pure speckle $\V \epsilon_k\in\mathbb{C}^{T}$, with no correlation along the spatial and the temporal axis ($\V \epsilon_k\sim \mathcal{N}_c(\M I)$), a multiplication by the matrix $\M L_k$, where $\M L_k\M L_k\T=\M \Gamma_k$ (e.g., $\M L_k$ is a Cholesky factor of coherence matrix $\M \Gamma_k$), gives a random vector that follows the distribution $\mathcal{N}_c(\M \Gamma_k)$. Thus, the speckle component can be generated from $\V \epsilon_k$ (Figure \ref{fig:GM}(b)):
\begin{align}
  \V s(\cdot,k)=\text{diag}(\sqrt{\V{r}(\cdot,k)})\M L_k\V \epsilon_k\,.
\end{align}
The vector $\V s\in\mathbb{C}^{TN}$ that concatenates all $T$ images one after another can be obtained by
\begin{align}
    \V s=
    \begin{pmatrix}
    \V s(t_1,\cdot)\\
    \vdots\\
    \V s(t_T,\cdot)
    \end{pmatrix} = 
    \text{diag}(\sqrt{\V{r}})
    \underbrace{
    \M{\Pi}^{-1}
    \begin{pmatrix}
    \M L_1&&\M0\\
    &\ddots\\
    \M 0& &\M L_N
    \end{pmatrix}
    \M{\Pi}}_{\M L}\V \epsilon\,.
\end{align}
The covariance matrix of the \emph{speckle component} $\V s$ is block diagonal after a proper permutation
\begin{align}
    \text{Cov}[\V s]= 
    \text{diag}(\sqrt{\V{r}})
    \M{\Pi}^{-1}\!\!
    \begin{pmatrix}
    \M \Gamma_1&&\M0\\
    &\ddots\\
    \M 0& &\M \Gamma_N
    \end{pmatrix}\!
    \M{\Pi}
    \text{diag}(\sqrt{\V{r}})\,,
\end{align}
which shows that correlations are only along the temporal axis of the spatio-temporal stack.

The \emph{dominant scatterers component} $\V d\in\mathbb{C}^{TN}$ contains non zero values only at pixels with dominant scatterers. Such scatterers may appear or disappear at some point in the time series.

The SLC amplitudes of the scene $\V z$ then correspond to the superimposition of the two components: $\V z=\V s+\V d$, see Figure \ref{fig:GM}(c).  
We model the effects of the atmospheric phase, the topographic (and possibly displacement) phase of the speckle component \cite{bamler1998synthetic}, and the spectral response of the SAR system as follows (Figure \ref{fig:GM}(d))
\begin{align}
   \tilde{\V z} &=\begin{pmatrix}
    \tilde{\V z}(t_1,\cdot)\\
    \vdots\\
    \tilde{\V z}(t_T,\cdot)
    \end{pmatrix}\nonumber\\
    &= \begin{pmatrix}
    \M H_1\diag(\exp(j\V\varphi_1))&&\M 0\\
    &\ddots&\\
    \M 0&&\M H_T\diag(\exp(j\V\varphi_T))
    \end{pmatrix}\V z\,,
\end{align}
where $\tilde{\V z}$ is the complex amplitude that includes these effects, $\M H_t\in\mathbb{C}^{N\times N}$ is the SAR response for the $t$-th acquisition, and $\V\varphi_t=\V\varphi_{\text{atmo}_t}+\V\varphi_{\text{topo}_t}+\V\varphi_{\text{disp}_t}\in\mathbb{C}^{N}$ combining the different sources of phase modification. The spectral response of the SAR system is generally identical for all passes, up to a 2D shift due to angular discrepancies (incidence and possibly squint angle differences between acquisitions). 
Linear operators $\M H_t$, $1\leq t\leq T $, can thus be written $\M H_t=\diag(\exp(-j\V\psi_t))\M Q\diag(\exp(j\V\psi_t))$, where $\M Q\in\mathbb{R}^{N\times N}$ is the real-valued operator (in spatial domain), corresponding to a spectral response (in Fourier domain) that is symmetrical and centered on the 0 frequency (i.e., 0 Doppler), and the phase vector $\V\psi_t$ is the 2D ramp corresponding to this 2D shift in Fourier domain (accounting for the angular discrepancies at pass $t$). The complex amplitudes of the $t$-th pass can be rewritten
\begin{align}
   \tilde{\V z}_t = 
    \diag(\exp(-j\V\psi_t))\M Q\diag(\exp(j\V\varphi_t+j\V\psi_t))\V z_t\,.\label{eq:opSAR}
\end{align}
The linear operator $\M Q$ accounts for the spectral apodization introduced to reduce the sidelobes of strong scatterers and a possible over-sampling (0-padding in Fourier domain), both inducing a low-pass filtering effect on SAR images that does not depend on $t$.

Since the multi-temporal stack $\tilde{\V z}$ is generated from $\V \epsilon$ through a series of linear operations, $\tilde{\V z}$ is also Gaussian distributed with a mean equal to $\tilde{\V d }$, where for each date $t$ the subvector $\tilde{\V d }(t,\cdot)\in\mathbb{C}^N$ is equal to $\M H_t\diag(\exp(j\V\varphi_t))\V d_t$, the low-pass filtered dominant scatterers component, and a covariance given at the bottom of page \pageref{eq:covfull}.
\begin{figure*}[!b]
\hrule
\begin{multline}
\hspace*{-3ex}\text{Cov}[\tilde{\V z}]=
\begin{pmatrix}
    \diag(\exp(-j\V\psi_1))\M Q\diag(\exp(j\V\varphi_1+j\V\psi_1))&&\M 0\\
    &\ddots&\\
    \M 0&&\diag(\exp(-j\V\psi_T))\M Q\diag(\exp(j\V\varphi_T+j\V\psi_T))
    \end{pmatrix}\text{diag}(\sqrt{\V{r}})
    \M{\Pi}^{-1}\\\hspace*{-2ex}
    \begin{pmatrix}
    \M \Gamma_1&&\M0\\
    &\ddots\\
    \M 0& &\M \Gamma_N
    \end{pmatrix}\!
    \M{\Pi}
    \text{diag}(\sqrt{\V{r}})
\begin{pmatrix}
    \diag(\exp(-j\V\varphi_1-j\V\psi_1))\M Q \T\diag(\exp(j\V\psi_1))&&\M 0\\
    &\ddots&\\
    \M 0&&\diag(\exp(-j\V\varphi_T-j\V\psi_T))\M Q \T\diag(\exp(j\V\psi_T))
    \end{pmatrix}\!\!.\nonumber
\end{multline}\label{eq:covfull}
\end{figure*}
Complex values in $\tilde{\V z}$ are both spatially and temporally correlated.

\subsection{Achieving statistical independence of the real/imaginary component at date $t_{\text{ref}}$}
\label{sec:condindep}

The principle of the self-supervised training proposed in \cite{MERLIN}, called MERLIN, consists of splitting the real and imaginary components of a single-date SLC image and exploiting their statistical independence.
Two differents tasks can be considered when extending speckle reduction to multi-temporal stacks: (i) the multiple-input single-output (MISO) framework where multiple dates are provided in input but only a single image at a reference date $t_\text{ref}$ is restored; (ii) the multiple-input multiple-output (MIMO) framework that restores at once all the dates provided in the input multi-temporal stack. In the following, we follow the MISO approach depicted in Figure \ref{fig:ppe_entrainement} for two reasons:
\begin{itemize}
\item the requirement of \emph{statistical independence} with respect to the inputs of the network is easier to achieve when a single output is considered;
\item in order for a MIMO network to output very different images in case of large changes, several \emph{independent paths} must emerge within the network architecture, which requires a \emph{huge network capacity} (i.e., many parameters) \cite{havasi2021training} and a \emph{careful initialization} to avoid getting stuck in poor quality local minima during training (as observed in our preliminary experiments).
\end{itemize} 

In our MISO multi-temporal approach, we provide the network with the multi-temporal SLC stack of $T$ images where the real part (or imaginary part) of the reference date $t_\text{ref}$ is excluded. This excluded component is then used to supervise the training under the assumption that it is statistically independent from the inputs (where the reflectivities $\V r$ and dominant scatterers $\V d$ are considered deterministic and only the speckle $\V \epsilon$ is random). Two preprocessing steps are required to ensure this independence.

First, the shift of the SAR system response in the spectral domain at date $t_\text{ref}$ induces correlations between real and imaginary components at this date\footnote{as discussed in \cite{MERLIN},  the Hermitian symmetry of the SAR transfer function must be ensured. This may require additional steps (e.g., demodulation, truncation of the spectrum).}.
A simple pre-processing step can be applied to recenter the spectrum of
the image at the reference date around the 0 frequency by multiplication by the 2D phase ramp $\exp(j\V\psi_{t_\text{ref}})$. In order to preserve interferometric coherence, we apply the same spectral shift to all dates (so that the relative shift between Fourier spectra remains unchanged). We denote the centered complex amplitudes by $\centered{\V z}$, defined by
\begin{align}
   \forall t,\,\centered{\V z}(t,\cdot) &= \diag(\exp(j\V\psi_{t_\text{ref}}))\tilde{\V z}(t,\cdot)
\end{align}
where the phase ramp $\V\psi_{t_\text{ref}}$ required to recenter the spectrum can be estimated from the power spectrum of image $\tilde{\V z}({t_\text{ref}},\cdot)$. This leads to the following simplified expression at $t_\text{ref}$:
\begin{align}
   \centered{\V z}(t_{\text{ref}},\cdot) &=
\M Q\diag(\exp(j\V\varphi_{t_\text{ref}}+j\V\psi_{t_\text{ref}}))\V z_{t_\text{ref}}\,.\label{eq:opSARcent}
\end{align}

Second, a whitening step may be necessary to address the correlations along the temporal axis, depending both on the coherence matrices $\M \Gamma_k$ (modeling how temporal decorrelations affect the scene) and the shifts induced by the phases $\V\psi_t$ (modeling geometric decorrelation according to the interferometric baselines).
In the context of multi-temporal speckle filtering, the stronger the correlations along the temporal dimension, the less useful the additional images. It is therefore recommended to consider time series with sufficient temporal speckle decorrelation for which no whitening step is necessary, as illustrated by our results in section \ref{sec:expes}. If images are in interferometric configuration with a large coherence, a whitening step is required. We describe a specific procedure in Appendix \ref{sec:appWhite} and denote by $\whitened{\V z}$ the stack after this preprocessing step, i.e., with minimal correlations along the temporal dimension ($\whitened{\V z}=\centered{\V z}$ in the absence of whitening step). Assumption \ref{ass:indep} summarizes that temporal correlations have been suppressed by the preprocessing step:

\begin{assumption}
The preprocessed image $\whitened{\V z}_{\text{ref}}$ at date $t_\text{ref}$ is statistically independent of the images $\whitened{\V z}_t$ for all dates $t\neq t_\text{ref}$.
\label{ass:indep}
\end{assumption}

In our MISO framework, we will consider two sets of inputs (noted $\mathscr{E}_a$ and $\mathscr{E}_b$) that contain all images $\whitened{\V z}_t$ except for the imaginary part $\whitened{\V b}_{\text{ref}}\in\mathbb{R}^N$ (respectively the real part $\whitened{\V a}_{\text{ref}}\in\mathbb{R}^N$) of $\whitened{\V z}_{\text{ref}}$:
\begin{align}
    \mathscr{E}_a&=\{\whitened{\V a}_{\text{ref}}\}\cup\{\whitened{\V z}_t|t\neq t_\text{ref}\}\text{ and}\nonumber\\
    \mathscr{E}_b&=\{\whitened{\V b}_{\text{ref}}\}\cup\{\whitened{\V z}_t|t\neq t_\text{ref}\}.\nonumber
\end{align}
In the following proposition, we show that these inputs are independent from the component set aside. This independence will be key to train a network fed with the input set $\mathscr{E}_a$ (or $\mathscr{E}_b$) under the supervision of loss function involving the component $\whitened{\V b}_{\text{ref}}$ (resp. $\whitened{\V a}_{\text{ref}}$).

\begin{proposition}\label{prop:indep}
Under assumption \ref{ass:indep}, the input set $\mathscr{E}_a$ is statistically independent from the imaginary part $\whitened{\V b}_{\text{ref}}$ at date $t_{\text{ref}}$, and similarly the input set $\mathscr{E}_b$ is statistically independent from the real part $\whitened{\V a}_{\text{ref}}$.
\end{proposition}

\begin{proof}
Under assumption \ref{ass:indep}, the image $\whitened{\V z}_{\text{ref}}$ is independent from all other images $\whitened{\V z}_t$ with $t\neq t_{\text{ref}}$. It remains to prove that the real and imaginary parts at time $t_{\text{ref}}$ are independent. 
According to our generative model of Sec.\ref{sec:genemodel}, they can be expressed in terms of the speckle $\V\epsilon_\text{ref}$ and the dominant scatterers $\V d_\text{ref}$
\begin{align}
    \begin{pmatrix}
    \whitened{\V a_\text{ref}}\\
    \whitened{\V b_\text{ref}}
    \end{pmatrix}=
    \begin{pmatrix}
    \Re(\centered{\V d_\text{ref}})\\
    \Im(\centered{\V d_\text{ref}})
    \end{pmatrix}
    +
    \M M
    \begin{pmatrix}
    \Re(\V \epsilon_\text{ref})\\
    \Im(\V \epsilon_\text{ref})
    \end{pmatrix}\,,\label{eq:defab}
\end{align}
where
\begin{align}
\centered{\V d_\text{ref}} = \M Q\diag(\exp(j\V\varphi_\text{ref}+j\V\psi_\text{ref})){\V d_\text{ref}}
\end{align}
and
\begin{align}\nonumber
\M M=
    \begin{pmatrix}
    \M Q\diag\bigl(\cos(\V\alpha_\text{ref}) \sqrt{\V r_\text{ref}}\bigr)&
    -\M Q\diag\bigl(\sin(\V\alpha_\text{ref}) \sqrt{\V r_\text{ref}}\bigr)\\
    \M Q\diag\bigl(\sin(\V\alpha_\text{ref}) \sqrt{\V r_\text{ref}}\bigr)&
    \M Q\diag\bigl(\cos(\V\alpha_\text{ref}) \sqrt{\V r_\text{ref}}\bigr)
    \end{pmatrix}
\end{align}
with $\V\alpha_\text{ref} = \V \varphi_{\text{ref}} +  \V \psi_{\text{ref}}$ and where the square root as well as the multiplications between vector $\sqrt{\V r_\text{ref}}$ and the cosine and sine are all applied entry-wise.

Given that $\Re(\V \epsilon_\text{ref})$ and $\Im(\V \epsilon_\text{ref})$ are independent and identically distributed according to a Gaussian distribution $\mathcal{N}(\V 0,\tfrac{1}{2}\M I)$, the real-valued vector formed by the real and imaginary components is also distributed according to a Gaussian distribution:
\begin{align}
    \begin{pmatrix}
    \whitened{\V a}_{\text{ref}}\\
    \whitened{\V b}_{\text{ref}}
    \end{pmatrix}\sim
    \mathcal{N}\biggl(
     \begin{pmatrix}
    \Re(\centered{\V d}_{\text{ref}})\\
    \Im(\centered{\V d}_{\text{ref}})
    \end{pmatrix},
    \tfrac{1}{2}\M M\M M\T
    \biggr)\label{eq:gaussdistribab}
    \end{align}
    with 
    \begin{align}
    \M M\M M\T=\begin{pmatrix}
    \M Q\diag(\V r_{\text{ref}})\M Q\T & \M 0\\
    \M 0 & \M Q\diag(\V r_{\text{ref}})\M Q\T
    \end{pmatrix}.
    \end{align}
    This shows that $\whitened{\V a_\text{ref}}$ and $\whitened{\V b_\text{ref}}$ are both jointly Gaussian and decorrelated, and thus, independent.
\end{proof}

\subsection{Self-supervised training strategy}
\label{sec:training}

In \cite{MERLIN}, the following single-date loss function has been introduced:
\begin{align}
    \mathcal{L}_{\text{MERLIN}}(\V a,\V u) = \sum_k \frac{1}{2}\log u_k+\frac{a_k^2}{u_k}\,.
    \label{eq:lossMERLIN}
\end{align}
It was applied to train a network fed with the imaginary part $\V b$ of a single SLC image and supervised by the corresponding real part $\V a$ through $\mathcal{L}_{\text{MERLIN}}(\V a,\V u)$ (where $\V u$ represents the output of the network), or conversely by providing $\V a$ to the network and supervising with $\mathcal{L}_{\text{MERLIN}}(\V b,\V u)$. Assuming that $\V a$ and $\V b$ are statistically independent, the network was shown to learn how to estimate the reflectivities.

We extend this loss to our multi-temporal MISO framework by replacing $\V a$ with 
$\whitened{\V a}_{\text{ref}}$ and $\V b$ with 
$\whitened{\V b}_{\text{ref}}$. The parameters $\V\theta$ of our regression model $f_{\V \theta}$ (i.e., the deep neural network) can be learned by minimizing the following multi-temporal extension of the MERLIN loss function:
\begin{multline}
    \argmin_{\V \theta}\;\mathbb{E}_{
    \substack{
    \whitened{\V b}_{\text{ref}}|\V r,{\V d}\\
    \hspace*{.2ex}
    {\mathscr{E}_a}|\V r,{\V d}}
    } \left[\mathcal{L}_{\text{MERLIN}}\left(\whitened{\V b}_{\text{ref}},f_{\V \theta}(\mathscr{E}_a)\right)\right]\\+
    \mathbb{E}_{
    \substack{
    \whitened{\V a}_{\text{ref}}|\V r,{\V d}\\
    \hspace*{.2ex}
    {\mathscr{E}_b}|\V r,{\V d}}
    } \left[\mathcal{L}_{\text{MERLIN}}\left(\whitened{\V a}_{\text{ref}},f_{\V \theta}(\mathscr{E}_b)\right)\right].
    \label{eq:multiMERLIN}
\end{multline}
According to Proposition \ref{prop:indep}, 
the inputs of the network $\mathscr{E}_a$ or $\mathscr{E}_b$ are independent from the images $\whitened{\V b}_{\text{ref}}$ and $\whitened{\V a}_{\text{ref}}$ used in the loss. 
It is thus impossible for the network to predict the stochastic component in these images (the output $\V u=\V a$ would minimize equation (\ref{eq:lossMERLIN}) but cannot be obtained from the inputs).

In the following proposition, we consider the family of all possible models $f_{\V \theta}$ that map the input images to a single output image. We then discuss in the proof of Prop.\ref{prop:est} the special case of a sub-family of models corresponding to a given parameterization of the regression model $f_{\V \theta}$ (for example, a fixed neural network architecture).

\begin{proposition}
The expectation of the multi-temporal MERLIN loss function (\ref{eq:multiMERLIN}) is minimal with respect to the predictions $f_{\V \theta}(\mathscr{E}_a)$ and $f_{\V \theta}(\mathscr{E}_b)$ if and only if 
$f_{\V \theta}(\mathscr{E}_a)=\tilde{\V r}_{\text{ref}}+2\Im(\centered{\V d}_{\text{ref}})^2$ and 
$f_{\V \theta}(\mathscr{E}_b)=\tilde{\V r}_{\text{ref}}+2\Re(\centered{\V d}_{\text{ref}})^2$,
where
$\tilde{\V r}_{\text{ref}}$ is the diagonal of covariance matrix $\M Q\diag(\V r_{\text{ref}})\M Q\T$ and $\centered{\V d}_{\text{ref}}=\M Q\diag(\exp(j\V\varphi_{\text{ref}}+j\V\psi_{\text{ref}})){\V d}_{\text{ref}}$.
\label{prop:min}
\end{proposition}

\begin{proof}
We start by expressing the values of the two expectations that appear in equation (\ref{eq:multiMERLIN}). They involve terms of the form $\mathbb{E}[\sum_k \whitened{\V a}_{\text{ref}}(k)^2/\V u(k)]$ 
and $\mathbb{E}[\sum_k \whitened{\V b}_{\text{ref}}(k)^2/\V v(k)]$, 
where $\V u=f_{\V{\theta}}(\mathscr{E}_b)$ 
and $\V v=f_{\V{\theta}}(\mathscr{E}_a)$.
They can be rewritten $\mathbb{E}[\whitened{\V a}_{\text{ref}}\T\diag(1/\V u)\whitened{\V a}_{\text{ref}}]=\text{Tr}\{\diag(1/\V u)\mathbb{E}[\whitened{\V a}_{\text{ref}}\whitened{\V a}_{\text{ref}}\T]\}$ where $1/\V u$ denotes an entry-wise inversion. By marginalization of the Gaussian distribution defined in (\ref{eq:gaussdistribab}), we obtain $\mathbb{E}[\whitened{\V a}_{\text{ref}}\whitened{\V a}_{\text{ref}}\T]=\Re(\centered{\V d}_{\text{ref}})\Re(\centered{\V d}_{\text{ref}})\T+\tfrac{1}{2}\M Q\diag(\V r_{\text{ref}})\M Q\T$. Similarly,
$\mathbb{E}
[\whitened{\V b}_{\text{ref}}\T\diag(1/\V u)\whitened{\V b}_{\text{ref}}]=\text{Tr}\{\diag(1/\V u)\mathbb{E}[\whitened{\V b}_{\text{ref}}\whitened{\V b}_{\text{ref}}\T]\}$ with
$\mathbb{E}[\whitened{\V b}_{\text{ref}}\whitened{\V b}_{\text{ref}}\T]=\Im(\centered{\V d}_{\text{ref}})\Im(\centered{\V d}_{\text{ref}})\T+\tfrac{1}{2}\M Q\diag(\V r_{\text{ref}})\M Q\T$. This leads to:
\begin{align}
    \mathbb{E}_{
    \whitened{\V b}_{\text{ref}}|\V r,{\V d}
    }
    \left[\mathcal{L}_{\text{MERLIN}}(\whitened{\V a}_{\text{ref}},\V u)\right]&=\sum_{k} \frac{1}{2}\log \V u(k)\nonumber\\
    &\hspace*{2.1em}
    +\frac{\Re(\centered{\V d}_{\text{ref}}(k))^2
    +\tfrac{1}{2}\tilde{\V r}_{\text{ref}}(k)}{\V u(k)}\\
    \mathbb{E}_{
    \whitened{\V a}_{\text{ref}}|\V r,{\V d}
    }
    \left[\mathcal{L}_{\text{MERLIN}}(\whitened{\V b}_{\text{ref}},\V v)\right]&=\sum_{k} \frac{1}{2}\log \V v(k)\nonumber\\
    &\hspace*{2.1em}
    +\frac{\Im(\centered{\V d}_{\text{ref}}(k))^2
    +\tfrac{1}{2}\tilde{\V r}_{\text{ref}}(k)}{\V v(k)}.    
\end{align}
A necessary condition for the expectations to be minimal is:
\begin{multline}
    \frac{\partial}{\partial \V u(k)}\mathbb{E}_{\whitened{\V a}_{\text{ref}}|\V r,{\V d}} \left[\mathcal{L}_{\text{MERLIN}}(\whitened{\V a}_{\text{ref}},\V u)\right]=0\\
    \Rightarrow
    \V u(k)=\tilde{\V r}_{\text{ref}}(k)+2\Re(\centered{\V d}_{\text{ref}}(k))^2
    \label{eq:uopt}
\end{multline}
\begin{multline}
    \frac{\partial}{\partial \V v(k)}\mathbb{E}_{\whitened{\V b}_{\text{ref}}|\V r,\V d} \left[\mathcal{L}_{\text{MERLIN}}(\whitened{\V b}_{\text{ref}},\V v)\right]=0\\
    \Rightarrow
    \V v(k)=\tilde{\V r}_{\text{ref}}(k)+2\Im(\centered{\V d}_{\text{ref}}(k))^2.
    \label{eq:vopt}
\end{multline}
The second-order derivatives for the values of $\V u(k)$ and $\V v(k)$ given by equations (\ref{eq:uopt}) and (\ref{eq:vopt})
\begin{multline}
    \left.
    \frac{\partial^2 \mathbb{E}_{\whitened{\V a}_{\text{ref}}|\V r,{\V d}} \left[\mathcal{L}_{\text{MERLIN}}(\whitened{\V a}_{\text{ref}},\V u)\right]}{\partial \V u(k)^2}
    \right|
    _{\V u(k)=\tilde{\V r}_{\text{ref}}(k)+2\Re(\centered{\V d}_{\text{ref}}(k))^2}\\
    =\frac{1}{2(\tilde{\V r}_{\text{ref}}(k)+2\Re(\centered{\V d}_{\text{ref}}(k))^2)^2}
    \label{eq:uder2}
\end{multline}
\begin{multline}
    \left.
    \frac{\partial^2 \mathbb{E}_{\whitened{\V b}_{\text{ref}}|\V r,{\V d}} \left[\mathcal{L}_{\text{MERLIN}}(\whitened{\V b}_{\text{ref}},\V v)\right]}{\partial \V v(k)^2}
    \right|
    _{\V v(k)=\tilde{\V r}_{\text{ref}}(k)+2\Im(\centered{\V d}_{\text{ref}}(k))^2}\\
    =\frac{1}{2(\tilde{\V r}_{\text{ref}}(k)+2\Im(\centered{\V d}_{\text{ref}}(k))^2)^2}
    \label{eq:vder2}
\end{multline}
are both strictly positive, which shows that the values of $\V u(k)$ and $\V v(k)$ correspond to a minimum. Since the solution to equations (\ref{eq:uopt}) and (\ref{eq:vopt}) is unique, we have identified the only minimum of the objective function.
\end{proof}

\begin{proposition}
Minimization of the expectation of the multi-temporal MERLIN loss function leads to an unbiased estimator $[f_{\V \theta}(\mathscr{E}_a)+f_{\V \theta}(\mathscr{E}_b)]/2$ of the sum of the low-pass filtered reflectivities $\tilde{\V r}_{\text{ref}}$ and of the intensity of the low-pass filtered dominant scatterers $|\centered{\V d}_{\text{ref}}|^2$ at date $t_{\text{ref}}$, provided that $f_{\V\theta}$ is sufficiently expressive (e.g., a deep neural network with sufficient width).\label{prop:est}
\end{proposition}
\begin{proof}
Under the Universal Approximation Theorem for width-bounded ReLU networks \cite{lu2017expressive}, a network with sufficient width can be built to approximate an arbitrary (Lebesgue-integrable) function $f_{\V\theta}$. If less expressive estimators $f_{\V\theta}$ are considered (smaller networks, not fully-connected architectures, other estimators than deep neural networks), a bias may appear due to the reduced ability of the estimator to match the optimal output given in Proposition \ref{prop:min}. 

For a sufficiently expressive estimator producing the optimal output, according to Proposition \ref{prop:min}, the minimum of the expectation of the multi-temporal MERLIN loss function is reached for $f_{\V \theta}(\mathscr{E}_a)=\tilde{\V r}_{\text{ref}}+2\Im(\centered{\V d}_{\text{ref}})^2$ and 
$f_{\V \theta}(\mathscr{E}_b)=\tilde{\V r}_{\text{ref}}+2\Re(\centered{\V d}_{\text{ref}})^2$. The computation of the average concludes the proof:
\begin{align}
    \forall k,\,\frac{f_{\V \theta}(\mathscr{E}_a)(k)+f_{\V \theta}(\mathscr{E}_b)(k)}{2}=\tilde{\V r}_{\text{ref}}(k)+|\centered{\V d}_{\text{ref}}(k)|^2.
\end{align}
\end{proof}

\begin{figure*}[!p]
    \centering
    \includegraphics[width=\textwidth]{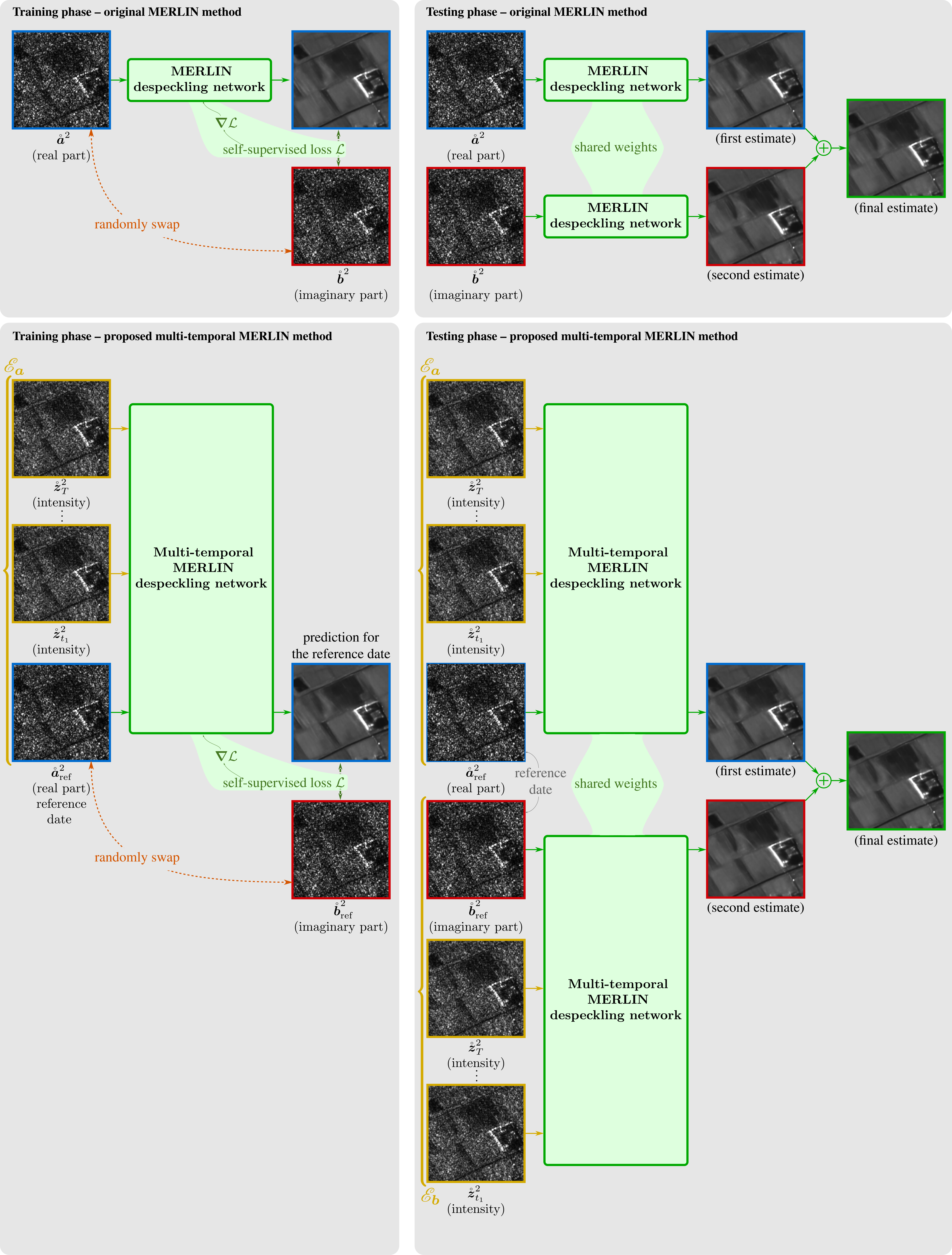}
    \caption{Principle of the self-supervised method MERLIN: original approach \cite{MERLIN} (top row) and proposed multi-temporal extension (bottom row).}
    \label{fig:ppe_entrainement}
\end{figure*}

Figure \ref{fig:ppe_entrainement} illustrates the principle of the proposed self-supervised training introduced in Propositions \ref{prop:min} and \ref{prop:est}: during training, we minimize MERLIN loss with the sets $\mathscr{E}_a$ or $\mathscr{E}_b$ as input and the images $\whitened{\V b}_{\text{ref}}$ or  $\whitened{\V a}_{\text{ref}}$ in the supervision. This leads to optimal weights $\V\theta^*$ at the end of the training phase. At test time, the estimates $f_{\V\theta^*}(\mathscr{E}_a)$ and $f_{\V\theta^*}(\mathscr{E}_b)$ are averaged to produce the final estimate.

For practical reasons, we use a convolutional U-Net architecture \cite{ronneberger2015u} (also used in the MERLIN method \cite{MERLIN})  with a small number of parameters, we consider a limited number of images in the training phase and an approximate minimization based on stochastic gradient computed over mini-batches. The estimator $f_{\V\theta^*}$ obtained is then only sub-optimal.

\section{Experiments}
\label{sec:expes}

The performance of the proposed multi-temporal MERLIN strategy is first studied on images with simulated speckle in paragraph \ref{sec:expes_simus}.
Results on Single Look Complex TerraSAR-X images are then presented in paragraph \ref{sec:expes_real}. In both cases, we compare multi-temporal MERLIN networks trained for an increasing number of additional inputs to study the quality improvement brought by these additional dates.
\subsection{Quantitative analysis on simulated speckle}
\label{sec:expes_simus}

The unsupervised learning strategy presented in Section \ref{sec:proposed_app} is motivated by the lack of speckle-free ground-truth images associated to each speckled SAR image. Yet, in order to perform a quantitative assessment of multi-temporal filtering, we first consider a simulated speckle framework in which both speckle-free and speckle-corrupted images are available. We build
high-quality speckle-free stacks by multi-temporal filtering with RABASAR-SAR2SAR \cite{RABASAR-SAR2SAR}. We then generate corrupted versions with simulated speckle corresponding to an ideal SAR transfer function, i.e., speckle with no spatial correlation in the simulated images. 
This reference data set is composed of 5 multi-temporal stacks of despeckled Sentinel-1 images, each stack containing from 25 to 69 images. Since the stacks are obtained from actual SAR images, realistic changes can be observed throughout the time series (e.g., evolution of the reflectivities in the fields). To simplify the simulations, we assume fully-developed speckle (the ground-truth images correspond to the reflectivities $\V r$ and no dominant scatterer is considered: $\V d=\V 0$). Information on the training sets and the hyperparameters used in all our network trainings are gathered in table \ref{table:hyperparameters}. The hyperparameters are kept unchanged whatever the number of additional inputs.

\begin{table}[!t]
    \centering
    \caption{Training parameters of the multi-temporal MERLIN networks (number of input channels from 2 to 20)}
\resizebox{\columnwidth}{!}{
\begin{tabular}{l@{}c@{\hspace*{5ex}} c}
\toprule
                                                                  & \bf Synthetic speckle                & \bf Actual speckle \\
                                                                  & \bf Sentinel-1                & \bf TerraSAR-X (Stripmap) \\                                                                  
\midrule
\bf \# stacks                                                         & 7                         & 2                         \\ 
\bf \# images                                                         & 237                         & 52                         \\ 
\bf avg images/stack          &          33.9        &  26\\                                                                         
\bf patch size                                                        & $256\times 256$           & $256\times 256$           \\ 
\bf batch size                                                        & 8                        & 8                        \\ 
\bf \# patches                                                        & 1616                     & 576		\\ 
\bf \# batches                                                        & 202                      & 72                      \\ 
\bf \# epochs                                                         & 1000                        & 1000                        \\ 
\multirow{3}{*}{\bf learning rate $\Biggl\{$}                                    & $10^{-3}$                 & $10^{-3}$                 \\ 
                                                                  & $10^{-4}$ after 10 epochs  & $10^{-4}$ after 10 epochs  			\\ 
                                                                  & $10^{-5}$ after 910 epochs & $10^{-5}$ after 910 epochs 			\\ 
                                                                  \bottomrule
\end{tabular}}

\label{table:hyperparameters}
\end{table}

\subsubsection{Impact of the number of additional channels}
\label{para:impact_additional_inputs}

We first evaluate the gain brought by the additional dates on the quality of the estimated speckle-free image. Depending on the presence or absence of change, including an additional input image may disturb or help the despeckling process.
When comparing the performances of two networks, a network with fewer inputs that underwent less changes might be favored over a network with more inputs which were all impacted by larger changes. We mitigate the impact of this phenomenon on our analysis by evaluating the performance of our networks on combinations of additional dates forming nested sets, i.e., a network with $j$ additional inputs, $j>i$, shares the same $i$ additional dates as a smaller network with $i$ additional inputs, but also benefits from $j-i$ supplementary inputs.

\begin{figure}[!t]
\centering
\includegraphics[width=\columnwidth]{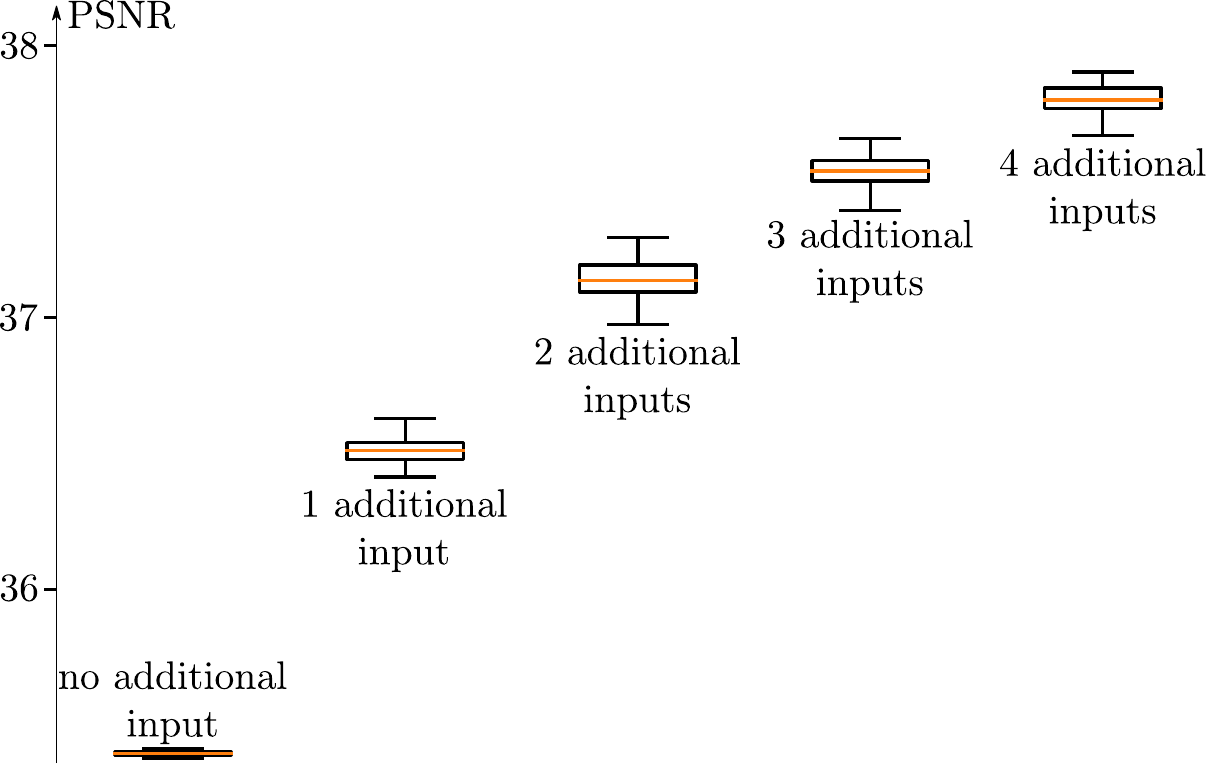}
\caption{Boxplots of PSNR values obtained for different draws of additional dates and various speckle realizations (each box plot indicates the minimum value, first quartile, in orange: median value, third quartile, and maximum).
}%
\label{fig:boxplots}
\end{figure}

Figure \ref{fig:boxplots} shows boxplots of the Peak Signal-to-Noise Ratio (PSNR) values computed on the log-reflectivities, for an increasing number of additional input images. The boxplots give for each configuration the minimum PSNR value; first, second, and third quartile PSNR values; and the maximum PSNR value. These statistics are computed over 88400 patches of $256\times 256$ pixels, corresponding to different spatial locations, choices of dates included as input, or speckle realizations. The restoration quality, measured by the PSNR values, improves with the number of images. This improvement is largest when the first additional dates are included, including a few more dates to an already large number of inputs produces a marginal improvement: unsurprisingly, multi-temporal filtering follows a law of diminishing returns with respect to the number of input dates.

Note that the dispersion of PSNR values for the mono-date filtering (leftmost boxplot of Figure \ref{fig:boxplots}) is very limited compared to the dispersion of PSNR values obtained with multi-temporal filtering. This is due to the variability of changes present in the additional channels: in multi-temporal filtering, situations with limited changes are more favorable to filtering and lead to better PSNR values while drawing a set of dates with larger changes inevitably gives a worse PSNR value (the variable luck in how similar the additional dates were explains the PSNR fluctuations).

\begin{figure*}[!t]
\centering
\includegraphics[width=\textwidth]{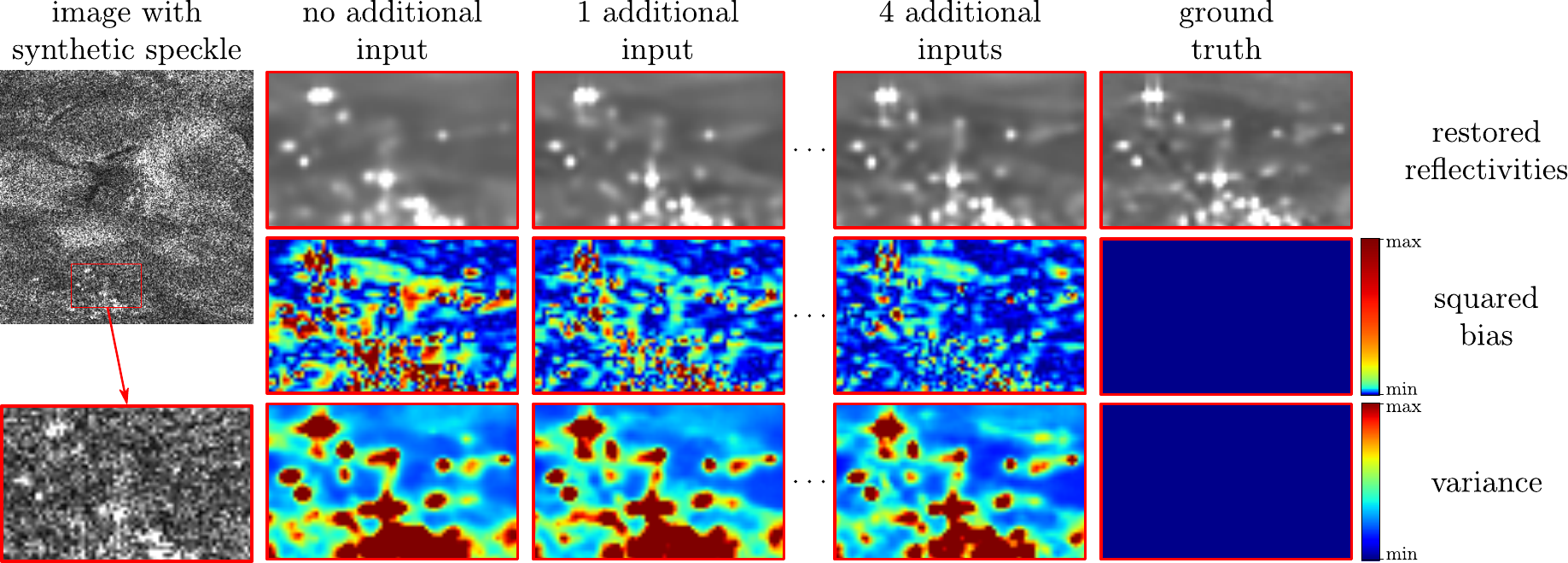}
\caption{Squared bias and variance averaged over 100 Multi-temporal MERLIN estimations of the reflectivities of a Sentinel-1 stack of Limagne (France). The speckle is simulated based on the method described in \cite{RABASAR-SAR2SAR}, and details of the test set are given in \ref{para:impact_additional_inputs}. }
\label{fig:error_maps}
\end{figure*}

As illustrated by Figure \ref{fig:error_maps}, PSNR values improve when increasing the number of additional input images due to the joint reduction of the estimation bias and of the estimation variance. Additional channels help preserve the spatial resolution, reducing the blur around sharp structures (such as points, lines, edges), as illustrated by the bias term. By not only combining spatial samples but also temporal samples, the estimation variance is reduced by multi-temporal filtering.

\begin{figure}[!t]
\centering
\includegraphics[width=\columnwidth]{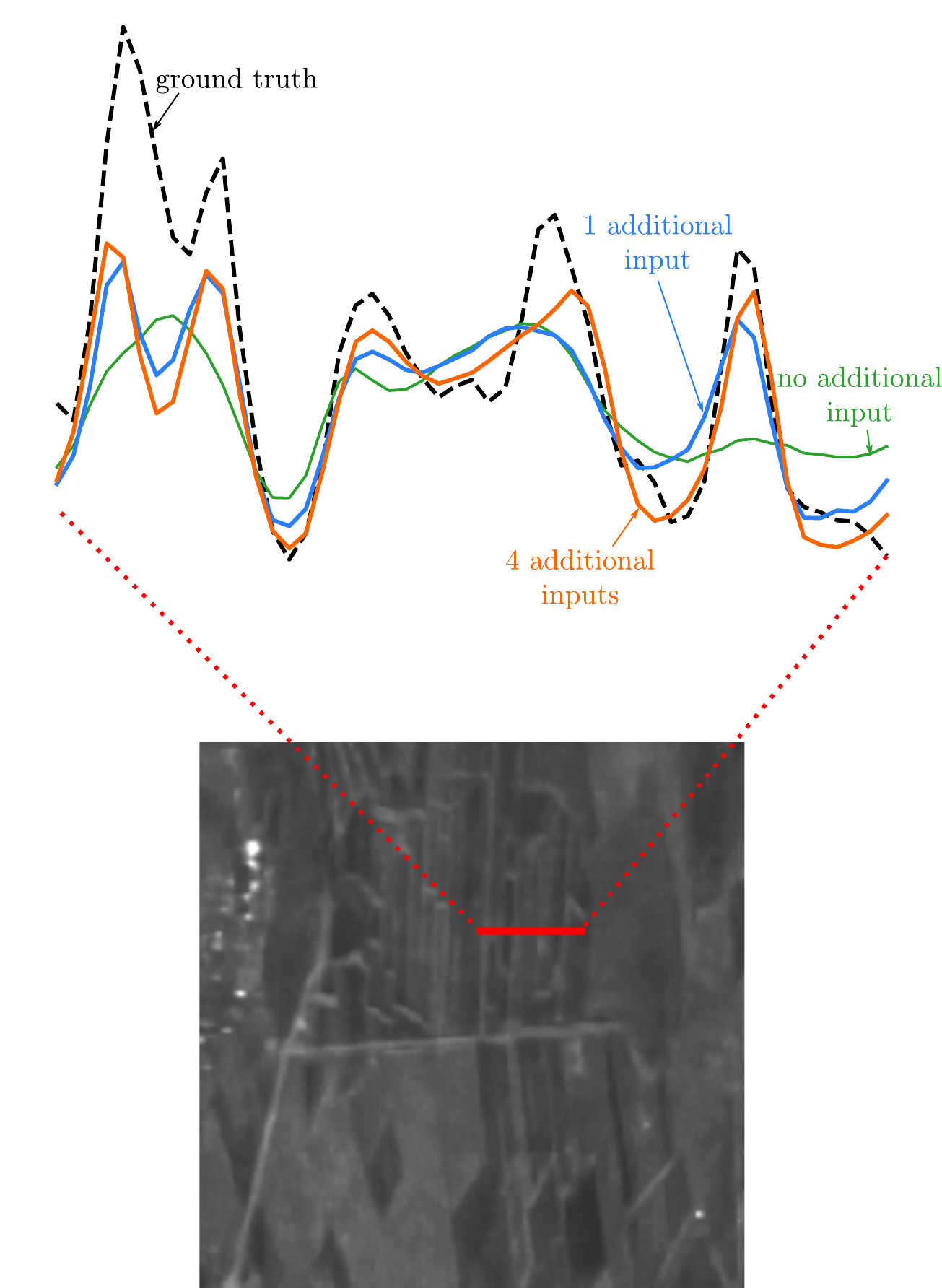}
\caption{Reflectivites profile along the red line, Marais1, date 14. The profile associated to MERLIN network estimation (green line) is blunt, meaning that the edges of the small observed structures are blurred. The more additional inputs there are, the sharper the profile lines, leading to a better retrieving of small structures. }
\label{fig:profile}
\end{figure}

The line profiles shown in Figure \ref{fig:profile} confirm the improved ability to restore fine structures with multi-temporal filtering (spatial resolution gain): processing a single date (green line) makes it difficult to retrieve the contrast of thin lines (hedges at the border of fields); with an additional date, or even better, with 4 additional dates, these structures are much better restored.

\subsubsection{Impact of temporal correlations}
\label{para:correlation_matters}

Images acquired in interferometric configuration may suffer from correlations along the temporal axis, as discussed in Section \ref{sec:proposed_app}. This is not ideal in the context of multi-temporal filtering as it reduces the potential benefit of temporal speckle averaging. Beyond this limitation, we illustrate here that, if neglected (i.e., if the temporal decorrelation step presented in Appendix \ref{sec:appWhite} is omitted), this type of correlations impacts the despeckling performance of networks trained with the multi-temporal MERLIN loss function (the independence assumption between the inputs and the component used for self-supervision is no longer valid).

We repeat the previous experiment with simulated speckle, this time introducing temporal correlations with a coherence matrix $\M\Gamma_k$ identical for all pixels $k$, and following a simple temporal decorrelation model
\begin{align}
    \forall k,\;\M\Gamma_k(t_i,t_j) = \text{exp} \left ( -\frac{|t_i - t_j|}{\tau}\right),
\end{align}
where $\tau$ is a characteristic decorrelation time. Rather than reporting how the despeckling performance degrades as a function of parameter $\tau$, we use the more intuitive average coherence $\Bar{\gamma}$ defined by
\begin{align}
    \Bar{\gamma} = \frac{1}{T^2}\!\sum_{1\leq i, j \leq T} \M\Gamma_k(t_i,t_j)\,. 
\end{align}

\begin{figure}[!t]
\centering
\includegraphics[width=\columnwidth]{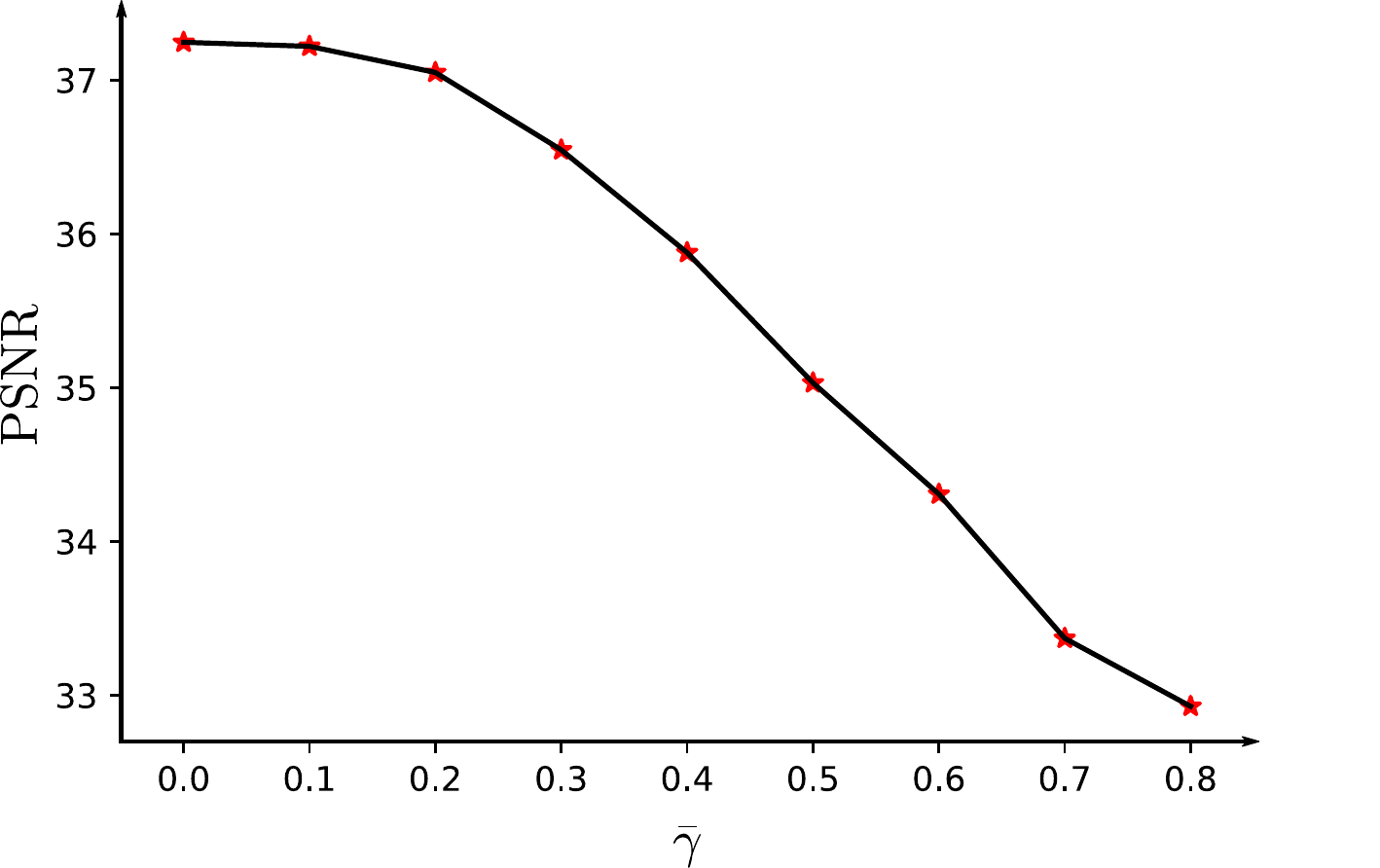}
\caption{Evolution of the restoration performance (PSNR values computed on log reflectivities) as a function of the average coherence $\Bar{\gamma}$ of the multi-temporal stack.}
\label{fig:psnr_correlation}
\end{figure}

Figure \ref{fig:psnr_correlation} reports the evolution of the PSNR of restored images (computed on log reflectivities) as a function of the average coherence $\Bar{\gamma}$ for a network that uses two additional inputs. Up to $\Bar{\gamma}\approx 0.2$ the performance is almost unchanged, then it degrades significantly. At $\Bar{\gamma}\approx 0.45$, the PSNR value is no better than that reached by a network with no additional input (mono-date filtering). Beyond $\Bar{\gamma}\approx 0.45$, it is worse to include additional dates. The reason is that the temporal correlations of speckle lead the network to "cheat" and to partially guess the speckled component in the images used to supervise the training. Once trained, the network systematically leaves a large fraction of the speckle fluctuations unchanged.

\subsection{Qualitative analysis of networks trained on actual SAR time series}
\label{sec:expes_real}

\begin{figure}[!t]
\centering
\includegraphics[width=\columnwidth]{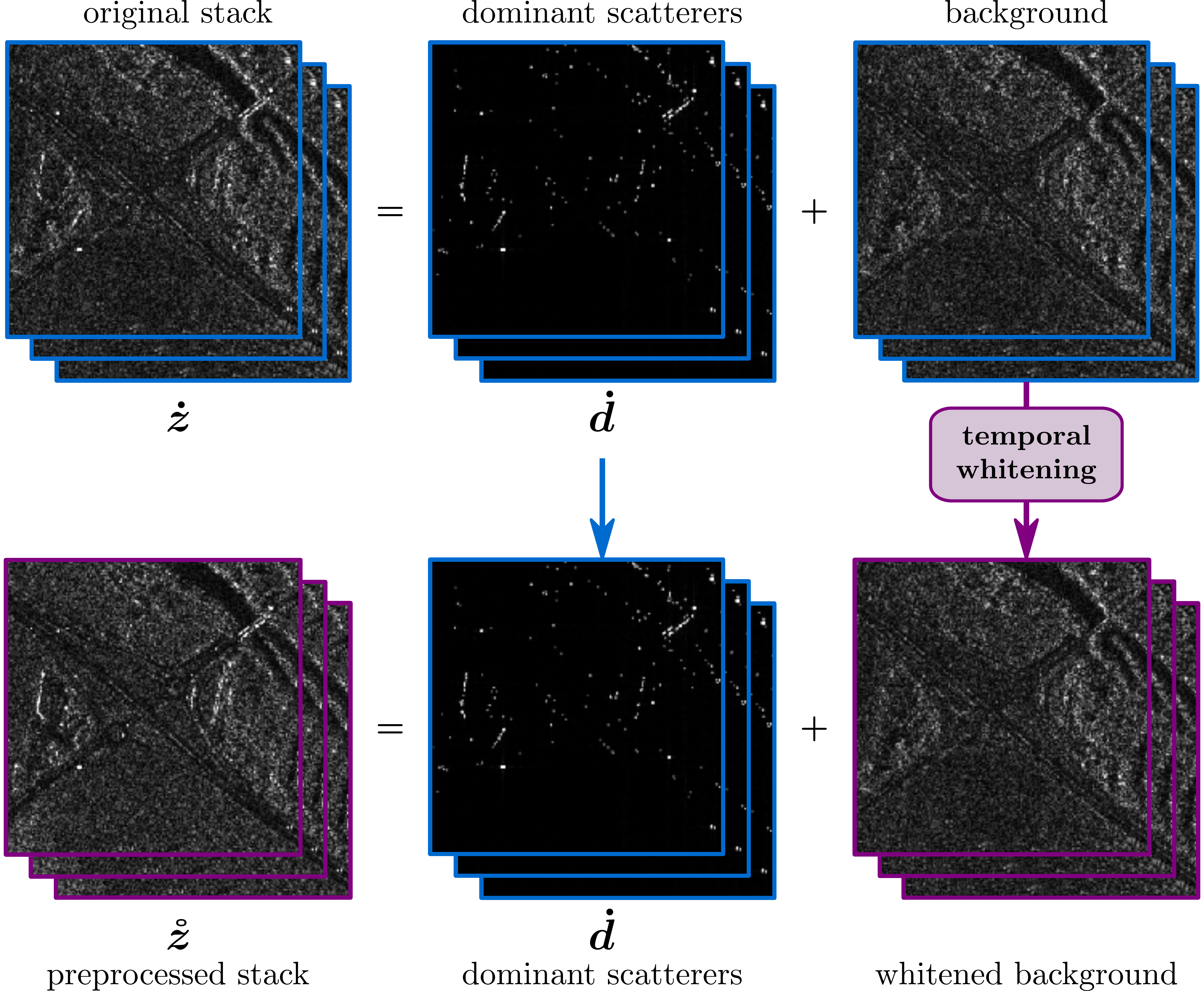}
\caption{Illustration of the preprocessing step to reduce temporal correlations of the speckle: (first row) the multi-temporal stack is decomposed into dominant scatterers and background using the method in \cite{abergel2018subpixellic}; (second row) the background component is then whitened and the dominant scatterers are added back to produce the preprocessed stack.}
\label{fig:remy}
\end{figure}

After the successful validation of our approach on times series with simulated speckle, we now turn to real speckle. 
First, we illustrate our (optional) preprocessing step that performs a temporal decorrelation with respect to the reference date. We recall here how this decorrelation is achieved, more details are given in \ref{sec:appWhite}:
\begin{enumerate}
    \item each SLC image of the stack is decomposed into a dominant scatterers component and a background component;
    \item interferograms with respect to the reference date are computed on the background components;
    \item at each pixel, a temporal whitening is performed based on the local coherence matrix estimated at the previous step;
    \item the contribution of dominant scatterers is reintroduced.
\end{enumerate}
Figure \ref{fig:remy} illustrates these different steps. We perform step 1) with the method described in \cite{abergel2018subpixellic}: the low-pass filtering effect introduced by the SAR system response (step d of the generative model of Figure \ref{fig:GM}) is first compensated by resampling and spectral equalization, then an \emph{a contrario} framework is applied to detect the cardinal sines of the dominant scatterers. The contribution of the dominant scatterers is then subtracted from the image and the original spectral apodization is reapplied. In step 2), we estimate interferograms between all pairs of images drawn from the stack of background components $\centered{\V z}-\centered{\V d}$. In our experiments, we use the MuLoG algorithm to compute these interferograms. This step is computationally intensive since forming all possible interferograms (in order to maximize the number of training samples to train our despeckling network) requires $\mathcal{O}(T^2)$ interferogram estimations for a multi-temporal stack with $T$ dates. Step 3) is much faster since it only requires applying pixelwise the simple whitening transform of equation (\ref{eq:white2canaux}). Finally, the reintroduction of dominant scatterers in step 4) leads to the temporally whitened stack $\whitened{\V z}$.

\begin{figure}[!t]
\centering
\includegraphics[width=\columnwidth]{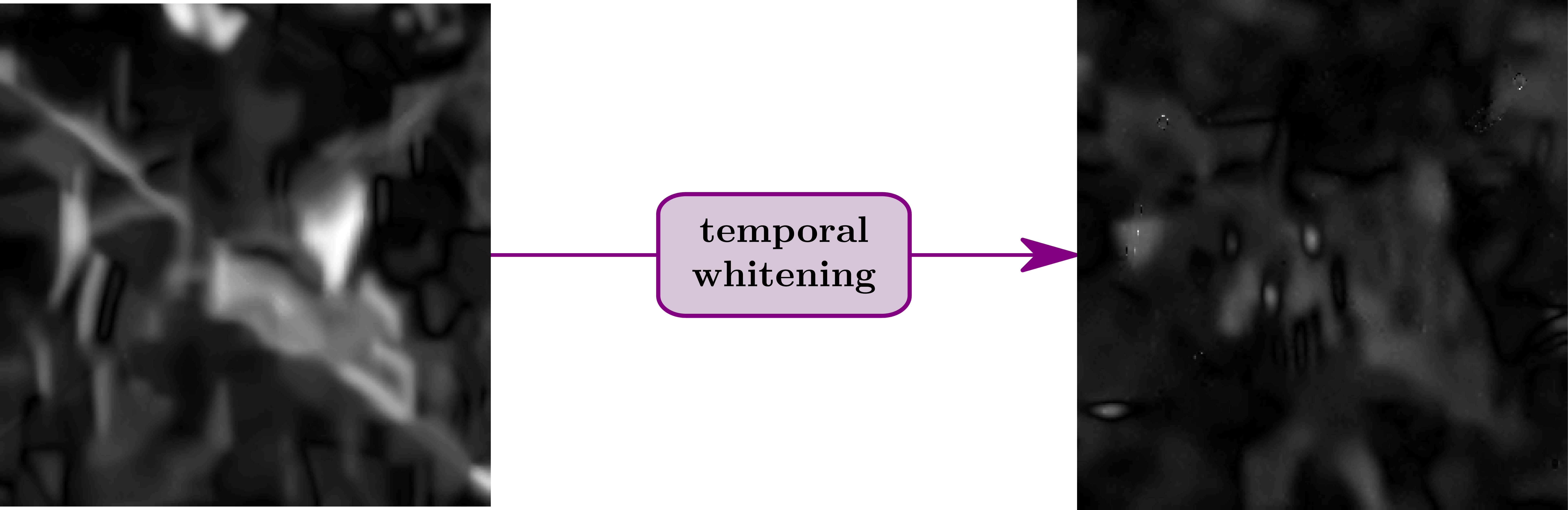}
\caption{Coherences computed with the MuLoG algorithm on the 2 first dates (2009/05/31 and 2009/06/11 ) of the Domancy TerraSAR-X stack \copyright DLR. The studied area is introduced in \ref{fig:remy}; left: estimated coherence before the whitening step; right: estimated coherence after the whitening  step.}
\label{fig:coherences}
\end{figure}

\begin{figure}[!t]
\centering
\includegraphics[width=\columnwidth]{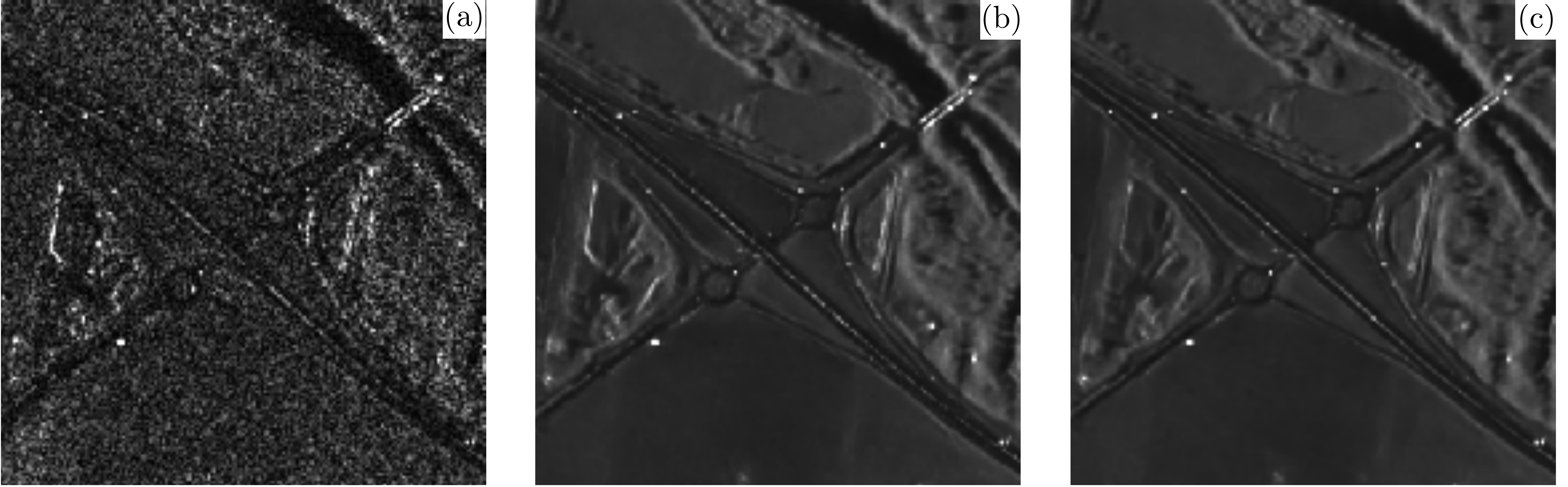}
\caption{Impact of the temporal whitening step on multi-temporal denoising of TerraSAR-X images \copyright DLR, city of Domancy (France). (a) noisy image; (b) multi-temporal MERLIN with 2 additional inputs trained on one TerraSAR-X stack without the temporal whitening step; (c) multi-temporal MERLIN with 2 additional inputs trained on one TerraSAR-X stack with the temporal whitening step.}
\label{fig:results_remy}
\end{figure}

In order to assess the impact of this temporal decorrelation step, we compared the performance of the same network trained in one case directly on a stack of 26 TerraSAR-X images $\centered{\V z}$ (i.e., the spectra have been shifted to center the spectrum of the reference date, but no temporal decorrelation step has been carried out), and in the other case using a pre-processed stack with our spectrum centering plus the temporal decorrelation technique. 
Coherences between the first two images of the original and the pre-processed stacks computed with the MuLoG algorithm are presented in Figure \ref{fig:coherences}. It shows that the proposed whitening step strongly reduces the coherence.
Despeckling results are presented in Figure \ref{fig:results_remy} and very few differences can be observed (slight changes may be noted around some scatterers). The average coherence on this stack of TerraSAR-X images is equal to 0.23, which corresponds to a mild level of correlation with a negligible impact on the despeckling performance, as shown in our experiments with simulated speckle reported in Figure \ref{fig:psnr_correlation}.
This illustrates that, even in the case of a satellite with interferometric capabilities, the computationally heavy preprocessing step of temporal decorrelation can be skipped when the coherence level is moderate.

\begin{figure*}[!t]
    \centering
    \includegraphics[width=\columnwidth]{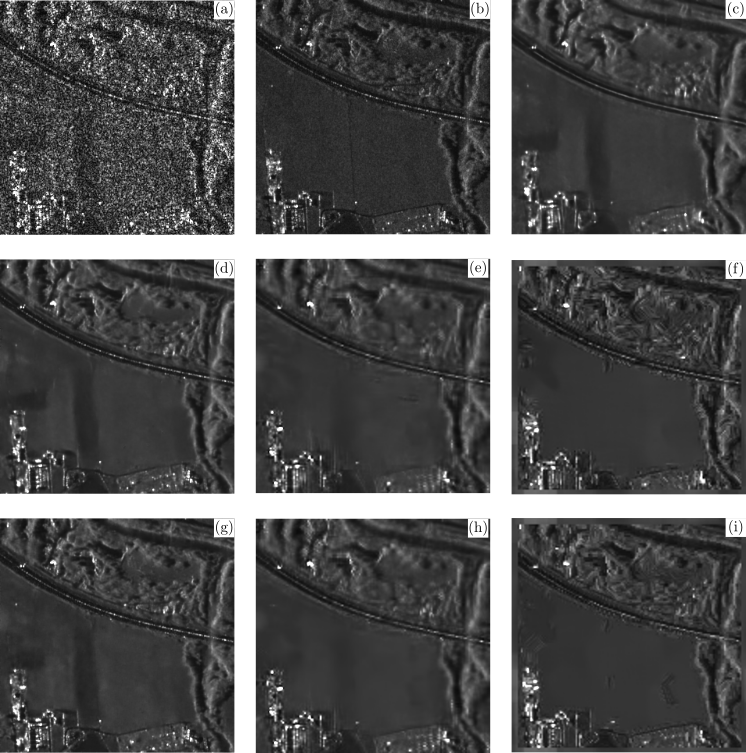}\;\;\;
    \includegraphics[width=\columnwidth]{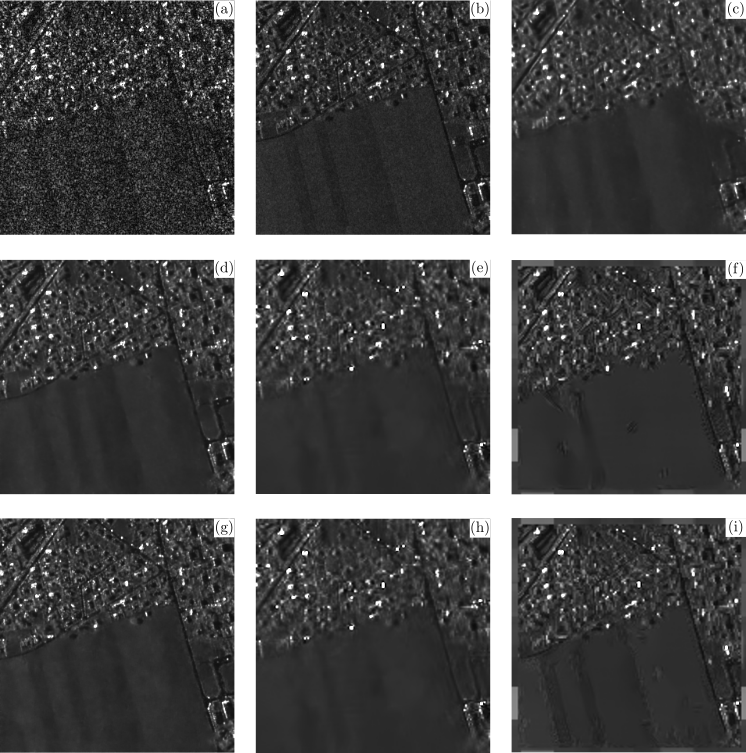}
    \caption{Multi-temporal denoising of TerraSAR-X images \copyright DLR (SLC images with actual speckle): left panel, city of Saint-Gervais (France); right panel, city of Domancy (France). Each panel shows (a) the noisy image; (b) the temporal average of all 26 images of the stack; (c) mono-date MERLIN filtering \cite{MERLIN}; (d) multi-temporal MERLIN with 3 additional inputs; (e) MSAR-BM3D \cite{chierchia2017sar} with 3 additional inputs, (f) 2SPPB with 3 additional inputs \cite{su2014two}; (g) multi-temporal MERLIN with 15 additional inputs; (h) MSAR-BM3D \cite{chierchia2017sar} with 15 additional inputs, (i) 2SPPB with 15 additional inputs \cite{su2014two}.}
    \label{fig:domancy_and_gervais_denoised}
\end{figure*}

Given the limited impact of this temporal whitening step for the multi-temporal stack we considered, we chose to skip this step and compare the performance of our network trained directly on multi-temporal stacks with other reference methods. Parameters used for our training are recalled in Table \ref{table:hyperparameters}, last column. 
Figure \ref{fig:domancy_and_gervais_denoised} shows two excerpts taken from the TerraSAR-X stacks used for training. Note that, given our self-supervised training strategy, our network can be tested on the same dataset as used for training. When applying the network to other datasets, the performance might drop if the type of area differs significantly (e.g, training on urban areas and testing on mountainous regions) due to a poor generalization. A fine-tuning step on the data of interest using the self-supervised loss is then preferable. The figure \ref{fig:domancy_and_gervais_denoised} contains two panels with the same numbering, each corresponding to a different stack. The single-look amplitude is shown in (a). In order to identify low-contrasted structures and fine details, the temporal average computed over the whole stack is shown in (b). Due to the changes that occur throughout the time series, this image is not directly comparable to image (a) but is still useful to analyze temporally-stable structures present in the scene given that speckle is strongly reduced by temporal averaging.  Areas with fluctuating reflectivities lead to an average value that differs from the reflectivity at the date of interest. Restoration results obtained with several speckle reduction methods are shown in each panel: (c) the mono-date MERLIN network, (d and g) the proposed multi-temporal MERLIN networks, and two baseline patch-based methods: (e and h) MSAR-BM3D introduced in \cite{chierchia2017sar} and (f and i) 2SPPB proposed in \cite{su2014two}. Multi-temporal methods are applied to a subset of 4 dates (the reference date + 3 additional dates) in the second row of the figure, or 16 dates (the reference date + 15 additional dates) on the last row. Temporal leakages can be observed in the results of MSAR-BM3D and 2SPPB: spurious information from the other dates contaminate the reference date, this is especially visible by the attenuation of the dark area (almost vertical rectangular field, in the center left of the image on the left pannel). In that respect, multi-temporal MERLIN offers much better results with restored reflectivities in good match with the noisy observation shown in Figure \ref{fig:domancy_and_gervais_denoised}(a). Edges are sharper and low-contrast structures are better preserved in the case with a limited amount of dates (3 additional inputs): Figure \ref{fig:domancy_and_gervais_denoised}(d-f) left and right panels.

\section{Conclusion}
\label{sec:conclu}
A generative model based on the decomposition of the SLC images into a speckle component and a dominant scatterers component has been introduced in this work. It breaks down the different sources of statistical correlation between spatial, temporal, and real/imaginary components of the complex amplitudes of SAR images. It shows that, under some assumptions like a low coherence or an adequate preprocessing step, the self-supervised training strategy MERLIN can be extended to stacks of SLC images. 

This strategy improves the despeckling performance achieved by mono-date networks by exploiting temporal redundancies of the scene and temporal fluctuations of speckle. Our quantitative analysis shows an improvement of the restored reflectivities, a refined spatial resolution, and very few temporal contamination by possible changes in the additional dates provided in input. Networks trained directly on SAR images, without groundtruth, produce restored images of higher quality compared to state-of-the-art techniques.

Deep neural networks trained with our self-supervised strategy seemingly bring a significant improvement to multi-temporal filtering in cases with limited or modest amounts of available dates. If numerous images are available, training a network to process all images becomes heavy, in particular regarding memory issues. Other approaches like ratio-based filtering \cite{RABASAR-SAR2SAR} or a different strategy to combine images from the stack may then be preferable.

The problem of image super-resolution in multi-spectral imaging has led to several multi-image fusion approaches based on deep learning \cite{DeepSUM,HighRes-net}. Future work may study whether the specific network architectures proposed in these methods would benefit the multi-temporal SAR despeckling problem.

Beyond multi-temporal filtering, our framework can straightforwardly be extended to multi-sensor or multi-modality fusion by including as additional input channels some images of the scene acquired by other sources.

\section*{Acknowledgments}
This project has been funded by the Futur \& Ruptures PhD program of the Fondation Mines-Telecom, and partially funded by ASTRAL project (ANR-21-ASTR-0011).

TerraSAR-X images were provided, as part of the project DLR-MTH0232 and DLR-LAN1746, by the German Space Agency DLR.

\appendices
\section{Pairwise temporal whitening}
\label{sec:appWhite}

Correlations along the temporal axis of the speckle depend both on the coherence matrices $\M \Gamma_k$ (capturing the temporal decorrelation of the scene) and on the shifts induced by the phases $\V\psi_t$ (accounting for the geometrical decorrelation due to the change of incidence angles introduced by the interferometric baseline). To reduce these correlations, a whitening process can be designed based on
the covariance values $\text{Cov}[\centered{\V z}(t_{\text{ref}},k);\,\centered{\V z}(t_i,k)]=\mathbb{E}[(\centered{\V z}_{\text{ref}}(k)-\centered{\V d}_{\text{ref}}(k)) (\centered{\V z}_{i}(k)-\centered{\V d}_{i}(k))^*]$.%

The dominant scatterer component $\centered{\V d}$ can be extracted from the images using an iterative algorithm \cite{abergel2018subpixellic}. The $2\times 2$ interferometric covariance matrices
$\begin{pmatrix}
\text{Cov}[\centered{\V z}(t_{\text{ref}},k);\,\centered{\V z}(t_{\text{ref}},k)]&
\text{Cov}[\centered{\V z}(t_{\text{ref}},k);\,\centered{\V z}(t_i,k)]\\
\text{Cov}[\centered{\V z}(t_i,k);\,\centered{\V z}(t_{\text{ref}},k)]&
\text{Cov}[\centered{\V z}(t_i,k);\,\centered{\V z}(t_i,k)]
\end{pmatrix}$ at each pixel $k$ can then be estimated by using an algorithm such as MuLoG \cite{MuLoG}. We propose to use these estimations to approximate the $2N\times 2N$ covariance matrix 
\begin{align}
    \text{Cov}\left[
    \begin{pmatrix}
    \centered{\V z}_{i}\\
    \centered{\V z}_{\text{ref}}
    \end{pmatrix}
    \right]\approx
    \begin{pmatrix}
    \M D_{i\,i} & \M D_{\text{ref}\,i}\T\\
    \M D_{\text{ref}\,i} &  \M D_{\text{ref}\,\text{ref}}
    \end{pmatrix},
    \label{eq:covD}
\end{align}
where the four $N\times N$ blocks are diagonal. Neglecting off-diagonal values of the matrices $\M D_{i\,i}$ and $\M D_{\text{ref}\,\text{ref}}$ amounts to considering a limited spatial correlation length (SAR impulse response is close to a Dirac). Neglecting off-diagonal values of the matrices $\M D_{i\,\text{ref}}$ and $\M D_{\text{ref}\,i}$ is justified when the multi-temporal stack is in interferometric configuration: a shift by one or more pixels of the image $\centered{\V z}_{i}$ with respect to image $\centered{\V z}_{\text{ref}}$ drastically reduces the interferometric coherence (i.e., the diagonal of $\M D_{i\,\text{ref}}$ is dominant). 

From the expression of the covariance matrix $\text{Cov}[\tilde{\V z}]$ given at the bottom of page \pageref{eq:covfull} and the definition of $\centered{\V z}$ in equation (\ref{eq:opSARcent}), we can derive the exact covariances $\text{Cov}[\centered{\V z}_{i}]$ and $\text{Cov}[\centered{\V z}_{{\text{ref}}}]$ of the centered complex amplitudes of the considered pair of SAR images: $\text{Cov}[\centered{\V z}_{i}]=\M Q \diag(\V{r}_{i})\M Q\T$ (and $\text{Cov}[\centered{\V z}_{{\text{ref}}}]=\M Q \diag(\V{r}_{{\text{ref}}})\M Q\T$, respectively). We approximate this covariance matrix by its diagonal: $\text{Cov}[\centered{\V z}_{i}]\approx \M D_{i\,i}$ (and $\text{Cov}[\centered{\V z}_{{\text{ref}}}]\approx \M D_{{\text{ref}} {\text{ref}}}$) with $\tilde{\V r}_{i}$ the diagonal of matrix $\M Q \diag(\V{r}_{i})\M Q\T$ (and $\tilde{\V r}_{{\text{ref}}}$ the diagonal of matrix $\M Q \diag(\V{r}_{{\text{ref}}})\M Q\T$ respectively). These vectors correspond to a low-pass filtered version of the reflectivity maps, according to the SAR response $\M Q$.

The anti-diagonal blocks are approximated by $\M D_{\text{ref} \, i} = \text{diag} (\tilde{\gamma}_{i\,\text{ref}} \sqrt{\tilde{\V r}_{i} \tilde{\V r}_{{\text{ref}}}})$ 
where products between vectors are applied entry-wise, and $\tilde{\V\gamma}_{i\,\text{ref}}\in\mathbb{C}^N$ is the vector of complex-valued coherences between dates $t_i$ and $t_{\text{ref}}$ ($\forall k,\,0\leq|\tilde{\V\gamma}_{i\,\text{ref}}(k)|\leq 1$).

The covariance matrix of a pair of complex amplitudes at a pixel $k$ is finally given by:
\begin{multline}
    \hspace*{-2ex}\text{Cov}\left[\begin{pmatrix}
    \centered{\V z}(t_i,k)-\centered{\V d}(t_i,k)\\
    \centered{\V z}(t_{\text{ref}},k)-\centered{\V d}(t_{\text{ref}},k)
    \end{pmatrix}\right]=
    \text{Cov}\left[\begin{pmatrix}
    \centered{\V z}(t_i,k)\\
    \centered{\V z}(t_{\text{ref}},k)\\    
    \end{pmatrix}\right]\\=
    \begin{pmatrix}
    \tilde{\V r}(t_i,k) & \tilde{\gamma}_{i\,\text{ref}}(k)\sqrt{\tilde{\V r}(t_i,k)\tilde{\V r}(t_\text{ref},k)}\\
    \tilde{\gamma}_{i\,\text{ref}}^*(k)\sqrt{\tilde{\V r}(t_i,k)\tilde{\V r}(t_\text{ref},k)}&\tilde{\V r}(t_{\text{ref}},k)\\    
    \end{pmatrix}\!\!.\hspace*{-1.1em}\label{eq:correlmat}
\end{multline}

The covariance along the temporal dimension between the image of reference $\centered{\V z}_{\text{ref}}$ and the image at date $t_i$ $\centered{\V z}_{i}$ modeled by (\ref{eq:covD}) can be suppressed by multiplying each $2N$ vector $(\centered{\V z}_{i} - \centered{\V d}_{i}, \centered{\V z}_{\text{ref}} - \centered{\V d}_{\text{ref}})$
by a whitening matrix $\M W$, leading to the whitened pair of images $(\whitened{\V z}_{i}, \whitened{\V z}_{{\text{ref}}})$:

\begin{align}
    \begin{pmatrix}
    \whitened{\V z}_{i}\\
    \whitened{\V z}_{{\text{ref}}}\\  
    \end{pmatrix}= \M W
    \begin{pmatrix}
    \whitened{\V z}_{i} - \centered{\V d}_{i}\\
    \whitened{\V z}_{{\text{ref}}} - \centered{\V d}_{{\text{ref}}}\\ 
    \end{pmatrix} +
    \begin{pmatrix}
    \centered{\V d}_{i}\\
    \centered{\V d}_{{\text{ref}}}\\
    \end{pmatrix} \\\nonumber
\end{align}

with

\begin{align}
    \M W&=\M \Pi^{-1
    }\!\!
    \begin{pmatrix}
    \M W_1\T & &\M 0\\
    &\ddots\\
    \M 0 & &\M W_N\T
    \end{pmatrix}\!
    \M \Pi\\
    \intertext{and}
    \M W_k\M W_k\T&=
    \text{Cov}\!\left[\begin{pmatrix}
    \centered{\V z}(t_i,k)-\centered{\V d}(t_i,k)\\
    \centered{\V z}(t_{\text{ref}},k)-\centered{\V d}(t_{\text{ref}},k)
    \end{pmatrix}\right]^{-1}.\label{eq:Wk}
\end{align}

The matrices $\M W_k$ can be obtained by Cholesky factorization of the inverse of the covariance matrix given in equation (\ref{eq:Wk}).

The closed-form expression of the Cholesky factorization in equation (\ref{eq:Wk}) leads to a simple definition of the whitened pair

\begin{align}
 \begin{cases}
    \whitened{\V z}(t_i,k)=
    \tau\centered{\V z}(t_i,k)+(1-\tau)\centered{\V d}(t_i,k)\\
    \hspace*{6.8em}-\sqrt{\frac{\tilde{\V r}(t_i,k)}{\tilde{\V r}(t_\text{ref},k)}}\tau\tilde{\gamma}_{i\,\text{ref}}^*(k)(\centered{\V z}(t_{\text{ref}},k)-\centered{\V d}(t_{\text{ref}},k))\\
    \whitened{\V z}(t_{\text{ref}},k)=\centered{\V z}(t_{\text{ref}},k)\,,
 \end{cases}
    \label{eq:white2canaux}
\end{align}
where $\tau=1/\sqrt{1-|\tilde{\gamma}_{i\,\text{ref}}(k)|^2}$.
Note that only the complex amplitude $\whitened{\V z}_{t_i}$ is modified while $\whitened{\V z}_{\text{ref}}$ is left unchanged.
This whitening procedure can thus be repeated for all pairs $(t_i,t_{\text{ref}})$, with $1\leq t_i\leq T$ and $t_i\neq t_{\text{ref}}$, thereby producing a pre-processed stack in which images are all decorrelated with respect to the reference date $t_{\text{ref}}$ (used in the subsequent processing as the target date for the despeckling task) and the decorrelated images provide information for the self-supervised training. Only the statistical independence with respect to this target date matters for the validity of the self-supervision used in section \ref{sec:training}.

We prove here that the whitened pair $\left(\whitened{\V z}(t_i,k), \whitened{\V z}(t_{\text{ref}},k) \right)$ has indeed a diagonal covariance matrix. 

We can rewrite the whitened pair as follows:
\begin{multline}
    \begin{pmatrix}
    \whitened{\V z}(t_i,k)\\
    \whitened{\V z}(t_{\text{ref}},k)\\    
    \end{pmatrix}
    =
    \begin{pmatrix}
    \tau& -\sqrt{\frac{\tilde{\V r}(t_i, k)}{\tilde{\V r}(t_{\text{ref}}, k)}} \tau\tilde{\gamma}_{i\,\text{ref}}^*(k)\\
    0&1\\    
    \end{pmatrix} \\
    \begin{pmatrix}
    \centered{\V z}(t_i,k) - \centered{\V d}(t_i, k)\\
    \centered{\V z}(t_{\text{ref}},k) - \centered{\V d}(t_{\text{ref}}, k) \\    
    \end{pmatrix}
    +
    \begin{pmatrix}
    \centered{\V d}(t_i, k)\\
    \centered{\V d}(t_{\text{ref}}, k) \\    
    \end{pmatrix}.\label{eq:whitenedfinal}
\end{multline}

Since the centered dominant component is deterministic, it follows from equations (\ref{eq:whitenedfinal}) and (\ref{eq:correlmat}) that

\begin{align}
    \text{Cov}&\left[\begin{pmatrix}
    \whitened{\V z}(t_i,k)\\
    \whitened{\V z}(t_{\text{ref}},k)\\    
    \end{pmatrix}\right]=
    \text{Cov}\left[\begin{pmatrix}
    \whitened{\V z}(t_i,k) - \centered{\V d}(t_i,k)\\
    \whitened{\V z}(t_{\text{ref}},k) - \centered{\V d}(t_{\text{ref}},k)\\    
    \end{pmatrix}\right]\nonumber\\
    &= \begin{pmatrix}
    \tau& -\sqrt{\frac{\tilde{\V r}(t_i, k)}{\tilde{\V r}(t_{\text{ref}}, k)}} \tau\tilde{\gamma}_{i\,\text{ref}}^*(k)\\
    0&1\\    
    \end{pmatrix} \nonumber\\
    &\hspace*{8.3ex}\text{Cov}\left[\begin{pmatrix}
    \centered{\V z}(t_i,k)\\
    \centered{\V z}(t_{\text{ref}},k)\\    
    \end{pmatrix}\right]
    \begin{pmatrix}
    \tau& 0\\
    -\sqrt{\frac{\tilde{\V r}(t_i, k)}{\tilde{\V r}(t_{\text{ref}}, k)}} \tau\tilde{\gamma}_{i\,\text{ref}}(k)&1\\
    \end{pmatrix} \nonumber\\
    &= \begin{pmatrix}
    \tilde{\V r}(t_i, k) & 0 \\
    0 & \tilde{\V r}(t_{\text{ref}}, k)
    \end{pmatrix}. 
\end{align}

This proves that, for each pixel $k$, the two complex amplitudes are decorrelated. Since they are jointly Gaussian and decorrelated, they are statistically independent.

\bibliographystyle{IEEEtran}
\bibliography{multi_t_merlin}
\end{document}